\documentclass[10pt,twocolumn]{article}
\usepackage{libertine}
\pdfoutput=1

\usepackage{graphicx}

\usepackage{booktabs}

\usepackage{algorithm}
\usepackage{balance}

\usepackage[noend]{algpseudocode}
\usepackage{subcaption}
\DeclareGraphicsExtensions{.pdf,.jpeg,.png}

\usepackage{multirow}
\usepackage{enumerate}
\usepackage{amsfonts}
\usepackage{amsmath}
\usepackage{hyperref}

\begin{document}

\title{Graph Operator Modeling over Large Graph Datasets\\ (Extended Version)}
     
\author{
Tasos Bakogiannis$^1$, Ioannis Giannakopoulos$^1$, Dimitrios Tsoumakos$^2$ and Nectarios Koziris$^1$ \\
\begin{minipage}[htb]{.4\linewidth}		
		\centering
\small{$^1$ Computing Systems Laboratory}\\
\small{School of ECE, NTUA, Athens, Greece} \\
\small{\{abk, ggian, nkoziris\}@cslab.ece.ntua.gr}
\end{minipage}
\quad
\begin{minipage}[htb]{.4\linewidth}
		\centering
\small{$^2$ Department of Informatics} \\
\small{Ionian University, Corfu, Greece} \\
\small{dtsouma@ionio.gr}
\end{minipage}
}
\date{}

\newcommand{\csubsubsection}[2]{\textit{\sffamily{#1\quad#2.}}}
\newcommand{\eat}[1]{}
\newcommand{\shortheader}[1]{\noindent$\triangleright$ \textbf{#1}}
\newenvironment{myitemize}
{
    \begin{list}{\labelitemi}{\leftmargin=1em}
        \setlength{\topsep}{0pt}
        \setlength{\parskip}{0pt}
        \setlength{\partopsep}{0pt}
        \setlength{\parsep}{0pt}
        \setlength{\itemsep}{0pt}
}
{
    \end{list}
}

\maketitle

\thispagestyle{empty}
\pagestyle{empty}

\begin{abstract} 
As graph representations of data emerge in multiple domains, data analysts need
to be able to intelligently select among a magnitude of different data graphs
based on the effects different graph operators have on them.  Exhaustive
execution of an operator over the bulk of available data sources is impractical
due to the massive resources it requires.  Additionally, the same process would
have to be re-implemented whenever a different operator is considered. To
address this challenge, this work proposes an efficient graph operator modeling
methodology. Our novel approach focuses on the inputs themselves, utilizing
graph similarity to infer knowledge about input graphs. The modeled operator is
only executed for a small subset of the available graphs and its behavior is
approximated for the rest of the graphs using machine learning techniques.  Our
method is operator-agnostic, as the same similarity information can be reused
for modeling multiple graph operators. We also propose a family of similarity
measures based on the degree distribution that prove capable of producing high
quality estimations, comparable or even surpassing other much more costly,
state-of-the-art similarity measures.  Our evaluation over both real-world and
synthetic graphs indicates that our method achieves extremely accurate modeling
of many commonly encountered operators, managing massive speedups over a
brute-force alternative.
\end{abstract}

\section{Introduction}

Graph Analytics has been gaining an increasing amount of attention in recent
years. Driven by the surge in social and business graph data, graph analytics
is used to effectively tackle complex tasks in many areas such as
bioinformatics, social community analysis, traffic optimization, fraud
detection, etc. A diverse collection of graph operators exists
\cite{tandfonline.com:Costa2007SurveyMeasurements}, with functionality
typically including the computation of centrality measures, clustering metrics
or network statistics \cite{dl.acm.org:Brandes:2005:NAM:1062400}.

Yet, as Big Data technologies mature and evolve (with regular advances in Big
Graph Systems \cite{Yan:BigGraph}), emphasis is placed on areas not solely
related to data (i.e., graph) size.  A different type of challenge steadily
shifts attention to the actual content: In content-based analytics
\cite{GANDOMI2015137}, data from social media platforms is processed for
sense-making. Similarly, in \emph{content-sensitive} applications such as
recommendation systems, web advertising, credit analysis, etc., the quality of
the insights derived is mainly attributed to the input content. The plethora of
available sources for content-sensitive analytics tasks now creates an issue:
Data scientists need to decide which of the available datasets should be fed to
a given workflow independently, in order to maximize its impact.  Yet, as
modern analytics tasks have evolved into increasingly long and complex series
of diverse operators, evaluating the utility of immense numbers of inputs is
prohibitively expensive. This is notably true for graph operators, whose
computational cost has led to extensive research on approximation algorithms
(e.g.,
\cite{DBLP:journals/datamine/RiondatoK16,DBLP:journals/jgaa/EppsteinW04}).

As a motivating example, let us consider a dataset consisting of a very large
number of citation graphs.  We wish to identify the graphs that have the most
well-connected citations and contain highly cited papers. The clustering
coefficient \cite{dl.acm.org:Brandes:2005:NAM:1062400}, a good measure of
neighborhood connectivity, would have to be computed for all the graphs in the
dataset in order to allow the identification of the top-k such graphs. To
quantify the importance of each paper, we consider a centrality measure such as
betweenness centrality \cite{dl.acm.org:Brandes:2005:NAM:1062400}.
Consequently, we would have to compute the maximum betweenness centrality score
for each citation graph and combine the results with those obtained from the
analysis based on the clustering coefficient. Yet, this is a daunting task due
to the operators' complexity and the number of executions required. It is not
straight-forward how the different input graphs affect the  output of the
clustering coefficient or betweenness centrality metrics. For traditional
algorithms, performance is driven by algorithmic complexity usually tied to the
input size. In Big Data Analytics, such analyses cannot be intuitively deduced
\cite{Halevy:UnreasonableDataEff}.

The challenge this work tackles is thus the following: Given a graph analytics
operator and a large number of input graphs, can we reliably predict operator
output \emph{for every input graph} at low cost? How can we rank or tangibly
characterize input datasets relative to their effect on job execution? In this
work, we introduce a novel, \textit{operator-agnostic} dataset profiling
mechanism: Rather that executing the operator over each input separately, our
work assesses the relationship between the dataset's graphs. Based on graph
similarity, we infer knowledge about them. In our example, instead of
exhaustively computing the clustering coefficient, we calculate a similarity
matrix for our dataset, compute the clustering coefficient for a small subset
of graphs and utilize the similarity matrix to estimate its remaining values.
We may then compute the maximum betweenness centrality for also a small subset
of citation graphs and \emph{reuse} the already calculated similarity matrix to
estimate betweenness centrality scores for the rest of the graphs.

Our method is based on the intuition that, for a given graph operator, similar
graphs produce similar outputs. This intuition is solidly supported by the
existence of strong correlations between different graph operators
(\cite{DBLP:journals/nhm/JamakovicU08,Bounova2012,Hernndez2015ClassificationOG}).
Hence, by assuming a similarity measure that correlates to a set of operators,
we can use machine learning techniques to approximate their outcomes. Given a
graph dataset and an operator to model, our method utilizes a similarity
measure to compute the similarity matrix of the dataset, i.e., all-pairs
similarity scores between the graphs of the dataset. The given operator is then
run for a small subset of the dataset; using the similarity matrix and the
available operator outputs, we are able to approximate the operator for the
remaining graphs. To the best of our knowledge, this is the first effort to
predict graph operator output over very large numbers of available inputs. In
summary, we make the following contributions in this work:
\begin{myitemize} 
\item We propose a novel, similarity-based method to estimate graph operator
output for very large numbers of input graphs. The method shifts the complexity
of numerous graph computations to less expensive pairs of similarities. This
choice offers two major advantages: First, our scheme is
\emph{operator-agnostic}: The resulting similarity matrix can be reused by
different operators, amortizing its computation cost. As a result, the cost of
our method is ultimately dominated by the computation of that operator for a
small subset of the dataset. Second, the method is agnostic to the similarity
measure that is used.  This property gives us the ability to utilize or
arbitrarily combine different similarity measures.  
\item We introduce a family of similarity measures based on the degree
distribution with a gradual tradeoff between detail and computation complexity.
Despite their simplicity, they prove capable of producing high quality
estimations, comparable or even surpassing other more costly, state-of-the-art
similarity measures
(\cite{Nature:Schieber2017StructuralDissimilarities,DBLP:journals/jmlr/VishwanathanSKB10}).
\item We improve on the complexity of the similarity matrix computation by
providing an alternative to calculating all-pairs similarity scores. We
propose, instead, to initially cluster a given dataset to groups of similar
graphs using the inverse of the similarity measure as a distance metric. Then,
calculate all-pairs similarities for each cluster, assuming inter-cluster
similarity scores to equal zero. In our experimental evaluation we observe that
this approach can lead to up to $15\times$ speedup in similarity matrix
calculations while having little to no effect on modeling accuracy.  
\item We offer an open-source
implementation\footnote{https://github.com/giagiannis/data-profiler} of our
method and perform an extensive experimental evaluation using both synthetic
and real datasets.  Our results indicate that the similarity-based approach is
accurately modeling a variety of popular graph operators, with errors even
$<1\%$, sampling a mere $5\%$ of the graphs for execution. Amortizing the
similarity cost over six operators, modeling is sped up to $18\times$ compared
to exhaustive modeling. Our proposed similarity measures produce similar or
more accurate results compared to state-of-the-art similarity measures but run
more than $5$ orders of magnitude faster.  Finally, our analysis provides
insights on the connection between different operators and the respective
similarity functions, demonstrating the utility of similarity matrix
composition.
\end{myitemize}

\section{Methodology}\label{methodology}

In this section, we formulate the problem and describe the methodology along
with different aspects of the proposed solution. We start off with some basic
notation followed throughout the paper and a formal description of our method
and its complexity.


Let a graph $G$ be an ordered pair $G = (V, E)$ with $V$ being the set of
vertices and $E$ the set of edges of $G$, respectively. The degree of a vertex
$u \in V$, denoted by $d_G(u)$, is the number of edges of $G$ incident to $u$.
The degree distribution of a graph $G$, denoted by $P_G(k)$, expresses the
probability that a randomly selected vertex of $G$ has degree $k$. A dataset
$D$ is a set of $N$ simple, undirected graphs $D = \{G_1, G_2, ..., G_N\}$. We
define a graph operator to be a function $g \colon D \to \mathbb{R}$, mapping
an element of $D$ to a real number. In order to quantify the similarity between
two graphs $G_a, G_b \in D$ we use a graph similarity function $s \colon D
\times D \to \mathbb{R}$ with range within $[0, 1]$. For two graphs $G_a, G_b
\in D$, a similarity of $1$ implies that they are identical while a similarity
of $0$ the opposite.


Consequently, the problem we are addressing can be formally stated as follows:
Given a dataset of graphs $D$ and a graph operator $g$, without knowledge of
the range of $g$ given $D$, we wish to infer a function $\hat{g} \colon D \to
\mathbb{R}$ that approximates $g$. Additionally, we wish our approximation to
be both accurate (i.e., $|g-\hat{g}|<\epsilon$, for some small $\epsilon$) and
efficient (i.e., $O(\hat{g})< O(g)$).  We observe that, although this
formulation resembles a function approximation problem, the two additional
requirements mentioned differentiate it from a typical problem of this class.
In such problems, we have knowledge of the entire output space of $g$ for a
given dataset. Moreover, no complexity restrictions are posed. In this
formulation, our goal is to provide an accurate approximation of $g$, while
avoiding its exhaustive execution over the entire $D$.


\begin{figure}[htb!]
    \centering 
    \includegraphics[width=.8\linewidth]{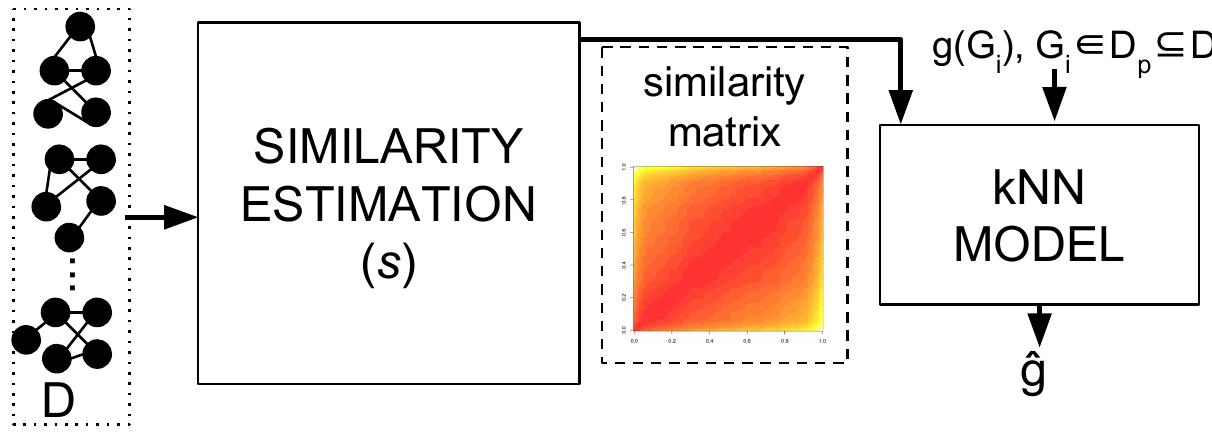}
    \caption{Methodology Pipeline}
    \label{fig:pipeline}
\end{figure}

To achieve this goal, we utilize the similarity matrix $R$, an $N \times N$
matrix with $R[i, j] = s(G_i, G_j)$, where $s$ is a given similarity measure.
As a result, $R$ contains all-pairs similarity scores between the graphs of
$D$.  $R$ is symmetric, its elements are in $[0, 1]$ and the entries of its
main diagonal equal $1$.  Our method takes as input a dataset $D$ and an
operator $g$ to model. It forms the pipeline depicted in Figure
\ref{fig:pipeline}: It begins with the computation of the similarity matrix $R$
based on $s$; it calculates the actual values of $g$ for a ratio $p \in (0,1)$
of randomly selected graphs of $D$, referred to as $D_p$.  Finally, it
estimates $g$ for the remaining graphs of $D$ by running a weighted version of
the $k$-Nearest-Neighbors (kNN) algorithm \cite{hastie2009elements}.  The
inferred function $\hat{g}$ is then given by the following equation:
\begin{equation}\label{eq:1} 
    \hat{g}(G_x) = \frac{\sum_{i \in \Gamma_k(x)} w_{xi} g(G_i)}{\sum_{i \in \Gamma_k(x)} w_{xi}} 
\end{equation} 
Where $w_{xi} = R[x, i]$ is the similarity score for graphs $G_x, G_i$, i.e.,
$w_{xi} = s(G_x, G_i)$, $\Gamma_k(x)$ is the set of the $k$ most similar graphs
to $G_x$ for which we have already calculated $g$ and $g(G_i)$ the value of the
operator for $G_i$. Our approach is formally described in Algorithm
\ref{modeling_algorithm}.

\begin{algorithm}
\caption{Graph Operators Modeling}\label{modeling_algorithm}
\begin{algorithmic}[1]
\Procedure{Approximate}{$[G_1, ..., G_N]$, $g$, $s$, $p$, $k$}
    \State $R \gets [\ ], T \gets \{\ \}, A \gets \{\ \}$
    \For{$(i, j) \gets [1, N] \times [1, N]$}
        \State $R[i, j] \gets s(G_i, G_j)$
    \EndFor

    \For{$i \gets 1, p \cdot N$}
        \State $r \gets randint(1, N)$
        \State $T[G_r] \gets g(G_r)$
    \EndFor

    \For{$x \gets [G_1, G_2, ..., G_N], x \notin keys(T)$}
        \State $t \gets findNeighbors(R, T, k, x)$
        \State $A[x] \gets calcApproximation(R, t)$
    \EndFor

    \State \textbf{return} $A$
\EndProcedure
\end{algorithmic}
\end{algorithm}

\setlength{\textfloatsep}{5pt}


The complexity of Algorithm \ref{modeling_algorithm} can be broken down to its
three main components. First, there is the calculation of the similarity matrix
$R$ in lines $3 - 4$, which, for a given similarity measure $s$ with complexity
$S$, runs in $O(N^2S)$. The second component (lines $5 - 7$), which computes
the operator $g$ for $pN$ graphs, has complexity $O(pNM)$, assuming that $g$
has complexity of $M$.  The approximation of the operator for the remaining
graphs (lines $8 - 10$) runs in $O((N(1 - p))((pN)log(pN) + k))$ since, for
each of the remaining dataset graphs (which are $N(1 - p)$), we first sort the
similarities of our training set ($T$) in order to find the $k$ nearest
neighbors to each unknown point ($findNeighbors$), an operation of
$O((pN)log(pN))$ complexity.  We then perform $k$ iterations to calculate the
weighted sum of Equ. \ref{eq:1} ($calcApproximation$). Thus, the complexity of
our method is:\\
\begin{equation} \label{eq:2}
    O(N^2S + pNM + (N(1 - p))((pN)log(pN) + k))
\end{equation}
From Equ. \ref{eq:2}, we deduce that the complexity of our method is dominated
by its first two components. Consequently, the lower the computational cost of
$s$, the more practical our approach will be.  Additionally, we expect our
training set to be much smaller than our original dataset (i.e., $p\ll 1$),
otherwise the second component will approach $NM$, which is no different that
calculating the operator for the entire dataset.

It is important to note here that the $O(N^2S)$ component corresponds to a
calculation performed only \textit{once}, whether modeling a single or multiple
operators. Indeed, as we show in this work, there exist both intuitive and
simple to compute operators that can be utilized for a number of different
graph tasks. Moreover, our methodology allows for \emph{composition} of
similarity matrices, permitting combinations of different measures. Given that
the similarity matrix calculation happens once per dataset, its cost gets
amortized over multiple graph operators, making the $O(pNM)$ factor the
dominant one for our pipeline.

\subsection{Similarity Measures}\label{methodology:similarity_measures}

The similarity matrix is an essential tool in our efforts to model graph
operators under the hypothesis that similar graphs produce similar operator
outputs. This would also suggest a connection between the similarity measure
and the graph operators we consider. Relative to graph analytics operators, we
propose a family of similarity measures based on graph degree distribution.
Reinforced by the proven correlations between many diverse graph operators
(\cite{DBLP:journals/nhm/JamakovicU08, Bounova2012,
Hernndez2015ClassificationOG}), we intend the proposed similarity measures to
express graph similarity in a way that enables modeling of multiple operators
at low cost.

\shortheader{Degree Distribution:} In order to quantify the similarity between
two graphs we rely on comparing their degree distributions. We compute the
degree distributions from the graph edge lists and compare them using the
Bhattacharyya coefficient BC \cite{Bhattacharyya1943Coefficient}. BC is
considered a highly advantageous method for comparing distributions
\cite{DBLP:journals/kybernetika/AherneTR98}.

BC divides the output space of a function into $m$ partitions and uses the
cardinality of each partition to create an $m$-dimensional vector representing
that space. As a measure of divergence between two distributions, the square of
the angle between the two vectors is considered. BC is thus calculated by:
$BC(q, r) = \sum_{i=1}^{m} \sqrt{q_i r_i}$, where $q$, $r$ are the two samples,
$m$ the number of partitions created by the algorithm and $q_i$, $r_i$ the
cardinality of the $i$-th partition of each sample.  In our case, the two
samples represent the graphs we intend to compare and the points in our space
are the degrees of the nodes of each graph. By dividing that space into
partitions and considering the cardinalities of each partition, we effectively
compare the degree distributions of those graphs.

In our implementation, we use a $k$-d tree \cite{DBLP:journals/cacm/Bentley75},
a data structure used for space partitioning to compute BC. We build a $k$-d
tree once, based on a predefined percentage of vertex degrees from all the
graphs in $D$. We then use the created space partitioning to compute degree
distributions for each graph. By using the \textit{median of medians} algorithm
\cite{Cormen:2009:IAT:1614191}, this process has complexity $O(\nu log\nu
+N|V|log\nu)$, where $\nu$ is the number of vertices used to build the tree.

\begin{figure}[htb!]
    \centering
    \includegraphics[width=.8\linewidth]{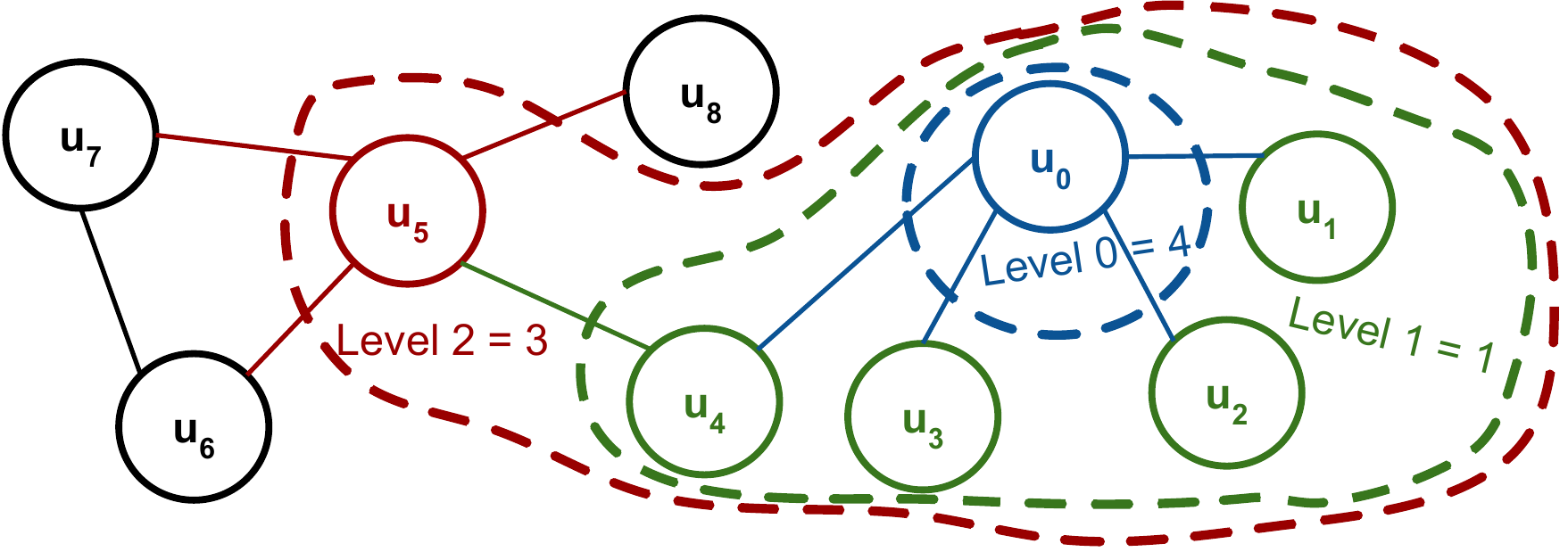}
    \caption{Example of Degree Distribution + Levels}
    \label{fig:distr_levels}
\end{figure}

\shortheader{Degree Distribution + Levels:} As an extension of the degree
distribution-based similarity measure, we consider a class of measures
with increasing levels of information. The intuition behind it is that the
degree of a vertex is a measure of its connectivity based on its immediate
neighbors. For instance, we can define the degree of a
vertex at level 1 as the degree of a supernode containing the vertex and
all its immediate neighbors. Internal edges of the supernode do not contribute
to his degree.  Generalizing this idea to more than one levels gives us a
measure of the indirect connectivity of a vertex. By combining the degrees of a
vertex for multiple levels we get information about its connectivity up to
\emph{level} hops away.  As an illustrative example, in Figure
\ref{fig:distr_levels} vertex \emph{$u_0$} has degree 4, when considering
its direct neighbors.  When its neighborhood is expanded to \emph{level} 1,
\emph{$u_0$}'s degree is 1 and for \emph{level} 2 it becomes 3.

Based on this idea, we quantify the similarity between graphs by calculating
the degrees up to a certain level for each vertex of the graphs and use BC to
compare the resulting degree distributions. A good property of this class of
measures is that they provide us with a nice tradeoff between accuracy and
computational cost. Increasing the number of degree distribution levels
involves additional computations but also incorporates more graph topological
insights to it. 
In order to calculate the degrees for a given level, for each vertex we perform
a depth-limited Depth First Search up to \emph{level} hops away in order to
mark the internal edges of the supernode. We then count the edges of the border
vertices (vertices \emph{level} hops away from the source) that do not connect
to any internal vertices. The complexity of this algorithm is
$\mathcal{O}(|V|(\overline{d})^l)$ where $\overline{d}$ is the average
branching factor (average degree) and $l$ the level-depth limit.

\shortheader{Degree Distribution + Vertex Count} A second extension to our
degree distribution-based similarity measure is based on the ability of our
method to combine similarity matrices. Graph size, in terms of vertex count, is
another graph attribute to measure similarity on. We formulate similarity in
terms of vertex count as: $s(G_i, G_j) = \frac{min(|V_{G_i}|,
|V_{G_j}|)}{max(|V_{G_i}|, |V_{G_j}|)}$. Intuitively, $s$ approaches $1$ when
$|V_{G_i}| - |V_{G_j}|$ approaches $0$, i.e., when $G_i, G_j$ have similar
vertex counts. To incorporate vertex count into the graph comparison, we can
combine the similarity matrices computed with degree distributions and vertex
counts using an arbitrary formula (e.g., linear composition).


\subsection{Similarity Matrix Computation Speedup}\label{methodology:scalability}

Based on Equ. \ref{eq:2}, the complexity of our method is primarily dominated
by the first two components and specifically by the similarity matrix
calculation. Although it refers to a \emph{one time} computation and its cost
is being amortized with modeling of multiple graph operators, having to compute
all-pairs similarity scores for a large collection of graphs can be
prohibitively expensive (in the order of $O(N^2)$ for $N$ graphs). Here, we
introduce a preprocessing step which we argue that improves the existing
computational cost, reducing the number of similarity calculations performed.

As, in order to approximate a graph operator, we employ kNN, we observe
that, for each graph, we only require the similarity scores to its
$k$ most similar graphs for which we have the value of $g$, i.e., the weights
in Equ. \ref{eq:1}. The remaining similarity scores involving this graph are
not used. Therefore we propose, as a first step, to run a clustering algorithm
which will produce clusters of graphs with high similarity. Then for each
cluster compute all-pairs similarity scores between its members. Inter-cluster
similarities are assumed to be zero and are not computed. By creating clusters
of size much larger than $k$, we expect minimal loss in accuracy while avoiding a
considerable number of similarity computations.

As a clustering algorithm we use a simplified version of $k$-medoids in
combination with $k$-means++, for the initial seed selection
(\cite{Kaufmann:ClusteringMedoids,DBLP:conf/soda/ArthurV07}). We aim at
creating clusters much larger than the required $k$ elements of kNN and
therefore do not consider necessary to recalculate the cluster medoids at each
iteration of the $k$-medoids algorithm.  In addition, we rely on $k$-means++ to
spread out the cluster medoids, something that has been proven to give better
clustering results \cite{DBLP:conf/soda/ArthurV07}.  Consequently, our approach
runs $k$-means++ to find $k$ cluster centers and groups the graphs based on
their distances to those centers.  Then, for each cluster computes all-pairs
similarity scores between its members.  As a distance measure we consider the
inverse of the similarity measure we employ, that is $d(G_i, G_j) = 1 - s(G_i,
G_j)$. Assuming the produced cluster sizes are close to $\frac{N}{c}$, $c$
being the number of clusters created, the similarity computations performed are
in the order of $O(\frac{N^2}{c} + Nc)$ .  Thus, by setting $c = \sqrt{N}$ we
can achieve $O(N\sqrt{N})$ similarity computations for a dataset of $N$ graphs.


\subsection{Discussion}\label{methodology:discussion}

In this section, we consider a series of issues that relate to the
configuration and performance of our method as well as to the relation between
modeled operators, similarity measure and input datasets.

\shortheader{Graph Operators:}\label{methodology:discussion:graph_operators} In
this work, we focus on graph analytics operators, namely centralities,
clustering metrics, network statistics, etc. Research on this area has
resulted in a large collection of operators, also referred to as \emph{topology
metrics}, with new ones being constantly added (e.g., \cite{tandfonline.com:Costa2007SurveyMeasurements,
dl.acm.org:Brandes:2005:NAM:1062400, Bounova2012,
Hernndez2015ClassificationOG}).  Topology metrics can be loosely classified in
three broad categories (\cite{DBLP:journals/nhm/JamakovicU08, Bounova2012,
Hernndez2015ClassificationOG}): Those related to \textit{distance},
\textit{connectivity} and \textit{spectrum}. In the first class, we find
metrics that involve distances between vertices such as \textit{diameter},
\textit{average distance} and \textit{betweenness centrality}. The second class
relates to vertex degrees containing metrics such as \textit{average degree},
\textit{degree distribution}, \textit{clustering coefficient}, etc.  Finally,
the third class comes from the spectral analysis of a graph and contains the
computation of \textit{eigenvalues}, the corresponding \textit{eigenvectors} or
other spectral-related metrics.\\
\shortheader{Similarity Matrix and Graph Operators:} An aspect we take under
consideration is the similarity measure we employ in relation to the graph
operator we intend to model.  Research has identified strong correlations
between certain classes of graph operators, for example topology metrics
(\cite{DBLP:journals/nhm/JamakovicU08, Bounova2012,
Hernndez2015ClassificationOG}). It is thus safe to assume that if, for example,
we base our similarity calculations on a degree-related measure, we should be
more successful in modeling degree-related graph operators than
distance-related. In practice, the analyst should have some level of intuition
on the type of graph operators to be modeled. In our experimental evaluation,
we model operators from all three aforementioned operator classes in order to
evaluate the efficiency of the proposed similarity measures.\\
\shortheader{Combining Similarity Measures:} We can think of use cases where we
want to quantify the similarity of graphs based on parameters unrelated to each
other which cannot be expressed by a single similarity measure. For example, we
might want to compare two graphs based on their degree distributions but also
take under account their order (vertex count).  Essentially, we would like to
be able to combine multiple independent similarity measures. This composition
can be naturally implemented in our system.  We can compute independent
similarity matrices for each of our similarity measures and 
``fuse'' those matrices into one based on a given formula. This technique is
presented in our experimental evaluation and proves effective in a number of
operators.\\
\shortheader{Regression Analysis} Although there exist several approaches to
statistical learning \cite{hastie2009elements}, we have opted for the kNN
method. We choose kNN for its simplicity and because we do not have to
calculate distances between points of our dataset (we already have that
information from the similarity matrix). The kNN algorithm is also suitable for
our use case since it is sensitive to localized data and insensitive to
outliers. A desired property, since we expect similar graphs to have similar
operator scores and should therefore be of influence in our estimations.
Conversely, we expect operator scores for graphs of low similarity to have
little or no influence on score estimations.

\section{Experimental Evaluation}\label{experimental_evaluation}

\subsection{Experimental Setup}

\shortheader{Datasets:} For our experimental evaluation, we consider both real
and synthetic datasets.  The real datasets comprise a set of ego graphs from
Twitter (\emph{TW}) which consists of 973 user ``circles'' as well as a dataset
containing 733 snapshots of the graph that is formed by considering the
Autonomous Systems (\emph{AS}) that comprise the Internet as nodes and adding
links between those systems that communicate to each other. Both datasets are
taken from the Stanford Large Network Dataset Collection \cite{SNAP:snapnets}.

We also experiment with a dataset of synthetic graphs (referred to as the
\emph{BA} dataset) generated using the SNAP library
\cite{SNAP:leskovec2016snap}.  We use the \texttt{{\fontfamily{pcr}\selectfont
GenPrefAttach}} generator to create random scale-free graphs with power-law
degree distributions using the Barabasi-Albert model
\cite{Science:Barabasi509}.  The degree distribution of the synthetic graphs,
according to this model, can be given by: $P(k) \sim k^{-\gamma}$,
where $\gamma = 3$.  We keep the vertex count of the graphs constant to 4K.  We
introduce randomness to this dataset by having the \emph{initial outdegree} of
each vertex be a uniformly random number in the range $[1, 32]$.  The
Barabasi-Albert model constructs a graph by adding one vertex at a time.  The
\emph{initial outdegree} of a vertex is the maximum number of vertices it
connects to, the moment it is added to the graph.  The graphs of the dataset
are simple and undirected.  Further details about the datasets can be found in
Table \ref{table:datasets}.

\shortheader{Similarity Measures:} \label{experimental_setup:similarity_measures}
We evaluate all the similarity measures proposed in Section
\ref{methodology:similarity_measures}, namely \emph{degree distribution +
levels}, for levels $0, 1, 2$ and \emph{degree distribution + vertex count}.
When combining vertex count with degree, we use the following simple formula:
$R = w_1 R_d + w_2 R_n$, with $R_d, R_n$ the degree distribution and vertex
count similarity matrices respectively. In our evaluation we set: $w_1=w_2=0.5$. 

To investigate their strengths and limitations, we compare them against two
measures functioning as our baselines. The first is a sophisticated
similarity measure not based on degree but rather on distance distributions
(from which the degree distribution can be deduced).  \textit{D-measure}
\cite{Nature:Schieber2017StructuralDissimilarities} is based on the concept of
network node dispersion ($\mathrm{NND}$) which is a measure of the
heterogeneity of a graph in terms of connectivity distances. 
From a computational perspective,
\textit{D-measure} is based on the all-pairs shortest paths algorithm, which
can be implemented in $O(|E| + |V|log(|V|))$ using Fibonacci heaps.  It is a
state-of-the-art graph similarity measure with very good experimental results
for both real and synthetic graphs.  It is considered efficient and since it
incorporates additional information to the degree distribution, it is suitable
to reason about how sufficient the measures we propose are.

Our second baseline comes from the extensively researched area of graph
kernels. Kernel methods for comparing graphs were first introduced in \cite{DBLP:conf/colt/GartnerFW03}.
Many kernels have been since proposed to address the problem of similarity in structured data
\cite{DBLP:journals/csr/GhoshDGQK18}. In our evaluation, we incorporate the
\textit{Random Walk Kernel} \cite{DBLP:conf/colt/GartnerFW03} which intuitively
performs random walks on a pair of graphs and counts the number of matching
walks as a measure of their similarity. For the purposes of our evaluation, we opted for
the geometric \textit{Random Walk Kernel} (\textit{rw-kernel}) as a widely used representative of
this class of similarity measures. The complexity of random walk
kernels is in the order of $O(|V|^6)$, however faster implementations with
speedups up to $O(|V|^2)$ exist \cite{DBLP:journals/jmlr/VishwanathanSKB10}.
In order to avoid the \textit{halting} phenomenon due to the kernel's decay
factor ($\lambda^k$) we set $\lambda = 0.1$ and the number of steps $k \leq 4$,
values that are considered to be reasonable for the general case
\cite{DBLP:conf/nips/SugiyamaB15}.  

\begin{table}[t!]
\small
\tabcolsep=0.1cm
\renewcommand{\arraystretch}{1.2}
\centering
\caption{Datasets overview}
\label{table:datasets}
\begin{tabular}{|c|c|c|c|l r|l r|}
\hline
\textbf{Name} &
\textbf{Size (N)} &
$\boldsymbol{\overline{|V|}}$ &
$\boldsymbol{\overline{|E|}}$ &
\multicolumn{2}{|c|}{\textbf{Range} $\boldsymbol{|V|}$} &
\multicolumn{2}{|c|}{\textbf{Range} $\boldsymbol{|E|}$} \\
\hline
\multirow{2}{*}{\textbf{TW}} &
\multirow{2}{*}{973} &
\multirow{2}{*}{132} &
\multirow{2}{*}{1,841} &
min: & 6 & min: & 9 \\
& & & & max: & 248 & max: & 12,387 \\
\hline
\multirow{2}{*}{\textbf{AS}} &
\multirow{2}{*}{733} &
\multirow{2}{*}{4,183} &
\multirow{2}{*}{8,540} &
min: & 103 & min: & 248 \\
& & & & max: & 6,474 & max: & 13,895 \\
\hline
\multirow{2}{*}{\textbf{BA}} &
\multirow{2}{*}{1,000} &
\multirow{2}{*}{4,000} &
\multirow{2}{*}{66,865} &
\multicolumn{2}{|c|}{\multirow{2}{*}{4,000}} &
min: & 3,999 \\ & & & & & & max: & 127,472 \\
\hline
\end{tabular}
\end{table}

\shortheader{Graph Operators:} In our evaluation, we model operators from all
the classes in Section \ref{methodology:discussion:graph_operators}. As
representatives of the distance class, we choose betweenness (\textbf{bc}),
edge betweenness (\textbf{ebc}) and closeness centralities (\textbf{cc}) (\cite{PhysRevE.69.026113,dl.acm.org:Brandes:2005:NAM:1062400}), three metrics
that express how central a vertex or edge is in a graph. The first two consider
the number of shortest paths passing from a vertex or edge while the third is
based on the distance between a vertex and all other vertices.  From the
spectrum class, we choose spectral radius (\textbf{sr}) and eigenvector
centrality (\textbf{ec}).  The first is defined as the largest eigenvalue of
the adjacency matrix of the graph. As a metric, it is associated with the
robustness of a network against the spreading of a virus \cite{ieee:4133799}.
The second is another measure that expresses vertex centrality
\cite{doi:10.1086/228631}. It is based on the eigenvectors of the adjacency
matrix. Finally, as a connectivity related metric we consider PageRank
(\textbf{pr}), a centrality measure used for ranking web pages based on
popularity \cite{DBLP:journals/cn/BrinP98}.

All measures, except spectral radius, are centrality measures expressed at
vertex level (edge level in the case of edge betweenness).  Since we wish all
our measures to be expressed at graph level, we will be using a method attributed
to Freeman \cite{Freeman1977Betweenness} to make that generalization. This is a
general approach that can be applied to any centrality
\cite{dl.acm.org:Brandes:2005:NAM:1062400}, and measures the average difference
in centrality between the most central point and all others: 
\begin{equation*}
c(G) = \frac{\sum_{i,j \in V} c(j)^{*} - c(i)}{|V| - 1} 
\end{equation*} 
 $c(G)$ being the measure at graph level, $c(i)$ the centrality value of the $i$-th vertex of $G$ and $c(j)^{*}$ the largest centrality value for all $i
\in V$.

All the graph operators are implemented in R. We use  the R package of the
\texttt{igraph} library \cite{R:csardi2006igraph} which contains
implementations of all the algorithms mentioned.

\shortheader{kNN:} The only parameter we will have to specify for kNN is $k$.
After extensive experimentation (omitted due to space constraints), we have
observed that small values of $k$ tend to perform better. As a result, all our
experiments are performed with $k = 3$.

\shortheader{Error Metrics:} \label{error_metrics} The modeling accuracy of our
method is quantified using two widely used measures from the literature, i.e.,
the \textit{Median Absolute Percentage Error} defined as: 
\begin{equation*} 
    MdAPE = median_{i=1,N}(100\frac{|g_i - \hat{g}_i|}{g_i}) 
\end{equation*}
where $g_i = g(G_i)$ is the $i\text{-}th$ actual value of a graph operator and
$\hat{g}_i$ the corresponding forecast. The second metric we use is the
\textit{Normalized Root Mean Squared Error}, defined as:
\begin{equation*}
    nRMSE = \frac{1}{c}\sqrt{\frac{1}{N}\sum_{i=1}^{N}(g_i - \hat{g}_i)^2}
\end{equation*}
with $c$ being the normalization factor which, in our case, is $max(g_i)$, $i
\in [1, N]$.  

\shortheader{Setup:} All experiments are conducted on an Openstack VM with 16
Intel Xeon E312 processors at 2GHz, 32G main memory running Ubuntu Server
16.04.3 LTS with Linux kernel 4.4.0. We implemented our prototype in Go
language (v.1.7.6).

\subsection{Experiments}

\begin{table*}[ht!]
\centering
\tabcolsep=0.1cm
\small
\caption{Modeling Errors and Execution Speedup for Different Sampling Rates}
\label{table:sampling_rate_vs_errors}
\begin{tabular}{|c c||r r r|r r r||r r r|r r r|}
\hline
\multirow{2}{*}{\textbf{Dataset}} & \multirow{2}{*}{\textbf{Metric}} &
\multicolumn{3}{|c|}{\textbf{MdAPE (\%)}} &
\multicolumn{3}{|c||}{\textbf{nRMSE}} &
\multicolumn{3}{|c|}{\textbf{Speedup $\times$}} &
\multicolumn{3}{|c|}{\textbf{Amortized Speedup $\times$}}
\\
\cline{3-14}
&
& \textbf{$p$=5\%} & \textbf{$p$=10\%} & \textbf{$p$=20\%}
& \textbf{$p$=5\%} & \textbf{$p$=10\%} & \textbf{$p$=20\%}
& \textbf{$p$=5\%} & \textbf{$p$=10\%} & \textbf{$p$=20\%}
& \textbf{$p$=5\%} & \textbf{$p$=10\%} & \textbf{$p$=20\%}
\\ \hline


\multirow{6}{*}{\textbf{AS}}
& \textbf{sr} &
1.3 & 1.1 & 0.9 &
0.05 & 0.03 & 0.02 &
6.4 & 3.8 & 3.3 &
\multirow{6}{*}{18.0} &
\multirow{6}{*}{9.5} &
\multirow{6}{*}{4.9}
\\
& \textbf{ec} &
0.1 & 0.1 & 0.0 &
0.01 & 0.00 & 0.00 &
5.7 & 4.5 & 3.1 &
\multicolumn{3}{|c|}{}
\\
& \textbf{bc} &
1.4 & 1.2 & 1.1 &
0.04 & 0.03 & 0.03 &
15.7 & 8.8 & 4.7 &
\multicolumn{3}{|c|}{}
\\
& \textbf{ebc} &
3.1 & 2.7 & 2.4 &
0.04 & 0.04 & 0.04 &
17.3 & 9.3 & 4.8 &
\multicolumn{3}{|c|}{}
\\
& \textbf{cc} &
0.4 & 0.4 & 0.3 &
0.01 & 0.01 & 0.01 &
14.0 & 8.2 & 4.5 &
\multicolumn{3}{|c|}{}
\\
& \textbf{pr} &
0.9 & 0.8 & 0.7 &
0.05 & 0.04 & 0.03 &
5.7 & 4.4 & 3.1 &
\multicolumn{3}{|c|}{}
\\ \hline


\multirow{6}{*}{\textbf{TW}}
& \textbf{sr} &
16.3 & 15.3 & 14.7 &
0.10 & 0.10 & 0.10 &
13.3 & 8.0 & 4.4 &
\multirow{6}{*}{14.8} &
\multirow{6}{*}{8.5} &
\multirow{6}{*}{4.6}
\\
& \textbf{ec} &
8.0 & 7.7 & 7.7 &
0.14 & 0.14 & 0.13 &
13.1 & 7.9 & 4.4 &
\multicolumn{3}{|c|}{}
\\
& \textbf{bc} &
17.8 & 17.5 & 16.8 &
0.16 & 0.15 & 0.14 &
13.0 & 7.8 & 4.4 &
\multicolumn{3}{|c|}{}
\\
& \textbf{ebc} &
29.5 & 29.8 & 28.6 &
0.12 & 0.12 & 0.12 &
13.5 & 8.0 & 4.4 &
\multicolumn{3}{|c|}{}
\\
& \textbf{cc} &
3.3 & 3.0 & 2.9 &
0.10 & 0.10 & 0.09 &
13.0 & 7.9 & 4.4 &
\multicolumn{3}{|c|}{}
\\
& \textbf{pr} &
9.2 & 7.7 & 7.2 &
0.07 & 0.06 & 0.05 &
13.2 & 7.9 & 4.4 &
\multicolumn{3}{|c|}{}
\\ \hline


\multirow{6}{*}{\textbf{BA}}
& \textbf{sr} &
3.3 & 1.8 & 0.9 &
0.04 & 0.03 & 0.03 &
5.6 & 4.4 & 3.0 &
\multirow{6}{*}{16.3} &
\multirow{6}{*}{9.0} &
\multirow{6}{*}{4.7}
\\
& \textbf{ec} &
0.4 & 0.3 & 0.3 &
0.01 & 0.01 & 0.01 &
3.7 & 3.1 & 2.4 &
\multicolumn{3}{|c|}{}
\\
& \textbf{bc} &
10.3 & 10.1 & 9.6 &
0.10 & 0.05 & 0.02 &
12.6 & 7.7 & 4.4 &
\multicolumn{3}{|c|}{}
\\
& \textbf{ebc} &
10.9 & 9.3 & 8.5 &
0.10 & 0.09 & 0.01 &
13.6 & 8.1 & 4.5 &
\multicolumn{3}{|c|}{}
\\
& \textbf{cc} &
2.4 & 2.2 & 2.1 &
0.04 & 0.04 & 0.03 &
9.9 & 6.6 & 4.0 &
\multicolumn{3}{|c|}{}
\\
& \textbf{pr} &
6.7 & 6.1 & 5.9 &
0.06 & 0.05 & 0.05 &
3.6 & 3.0 & 2.3 &
\multicolumn{3}{|c|}{}
\\ \hline

\end{tabular}
\end{table*}

\shortheader{Modeling Accuracy:} \label{experiments:modeling_accuracy} To
evaluate the accuracy of our approximations, we calculate \textit{MdAPE} and
\textit{nRMSE} for a randomized $20\%$ of our dataset: For a dataset of 1,000
graphs, $200$ will be chosen at random for which the error metrics will be
calculated. We vary the sampling ratio $p$, i.e., the number of graphs for
which we actually execute the operator, divided by the total number of graphs
in the dataset.  The results are displayed in Table
\ref{table:sampling_rate_vs_errors}. Each row represents a combination of a
dataset and a graph operator with the corresponding error values for different
values of $p$ between $5\%$ and $20\%$.

The results in Table \ref{table:sampling_rate_vs_errors} showcase that our
method is capable of modeling different classes of graph operators with very
good accuracy. Although our approach employs a degree distribution-based
similarity measure, we observe that the generated similarity matrix is
expressive enough to allow the accurate modeling of distance- and
spectrum-related metrics as well, achieving errors well below 10\% for most
cases. In \emph{AS} graphs, the $MdAPE$ error is less than $3.2\%$ for all the
considered operators when only a mere $5\%$ of the available graphs is
examined. Operators such as closeness or eigenvector centralities display low
$MdAPE$ errors in the range of $< 8\%$ for all datasets. Through the use of
more expressive or combined similarity measures, our method can improve on
these results,  as we show later in this Section.  We also note that the
approximation accuracy increases with the sampling ratio. This is expressed by
the decrease of both $MdAPE$ and $nRMSE$ when we increase the size of our
training set. These results verify that modeling such graph operators is not
only possible, but it can also produce highly accurate models with marginal
errors.

Specifically, in the case of the \emph{AS} dataset, we observe that all the
operators are modeled more accurately than in any other real or synthetic
dataset. This can be attributed to the topology of the \emph{AS} graphs.
These graphs display a linear relationship between vertex and edge counts.
Their clustering coefficient displays very little variance, suggesting that as
the graphs grow in size they keep the same topological structure. This gradual,
uniform evolution of the \emph{AS} graphs leads to easier modeling of the
values of a given graph topology metric.

On the other hand, our approach has better accuracy for degree- than
distance-related metrics in the cases of the \emph{TW} and \emph{BA} datasets.
The similarity measure we use is based on the degree distribution that is only
indirectly related to vertex distances. This can be seen, for example, in the
case of \emph{BA} if we compare the modeling error for the betweenness
centrality (bc) and PageRank (pr) measures.  Overall, we see that eigenvector
and closeness centralities are the two most accurately approximated metrics
across all datasets. After we find PageRank, spectral radius, betweenness and
edge betweenness centralities.

Willing to further examine the connection between modeling accuracy and the
type of similarity measure used, we have experimented with different
similarity measures, leading to the inclusion of \textit{D-measure} and
\textit{rw-kernel} in our evaluation. This has also lead to the development of the
degree-level similarity measures and the combination of similarity
matrices in the cases of degree distribution similarity matrix and vertex count
similarity matrix.

\shortheader{Execution Speedup:} Next, we evaluate the gains our method can
provide in execution time when compared to the running time of a graph operator
being executed for all the graphs of each dataset. Similarity matrix
computation is a time-consuming step that is executed once for each dataset.
Yet, an advantage of our scheme is that it can be reused for different graph
operators. Consequently, time costs can be amortized over different operators.
In order to provide a better insight, we calculate two types of speedups: One
that considers the similarity matrix construction from scratch for each
operator separately (provided in the \emph{Speedup} column of Table
\ref{table:sampling_rate_vs_errors}) and one that expresses the average speedup
for all six metrics for each dataset, where the similarity matrix has been
constructed only once (provided in the \emph{Amortized Speedup} column of Table
\ref{table:sampling_rate_vs_errors}).  For example, in the case of the
\emph{AS} dataset and for the spectral radius metric, our approach is $6.4
\times$ faster when using 5\% sampling ratio than the computation of the
spectral radius for all the graphs of \emph{AS}. Additionally, if we utilize
the same matrix for all six operators, this increases the speedup to
18$\times$. 

The observed results highlight that our methodology is not only capable of
providing models of high quality, but also does so in a time-efficient manner.
A closer examination of the Speedup columns shows that our method is
particularly efficient for complex metrics that require more computation time
(as in the \emph{ebc} and \emph{cc} cases for all datasets).  The upper bound
of the theoretically anticipated speedup equals $\frac{1}{p}$, i.e., in the
$p=5\%$ case each operator runs on 20 times fewer graphs than the exhaustive
modeling, without taking into account the time required for the similarity
matrix and the training of the kNN model. Interestingly, the \emph{Amortized
Speedup} column indicates that when the procedure of constructing the
similarity matrix is amortized to the six operators under consideration, the
achieved speedup is very close to the theoretical one. This is indeed the case
for the \emph{AS} and \emph{BA} datasets that comprise the largest graphs, in
terms of number of vertices: For all $p$ values, the amortized speedup closely
approximates $\frac{1}{p}$. In the case of the \emph{TW} dataset which consists
of much smaller graphs and, hence, the time dedicated to the similarity matrix
estimation is relatively larger than the previous cases, we observe that
the achieved speedup is also sizable. In any case, the capability of reusing
the similarity matrix, which is calculated on a per-dataset rather than on a
per-operator basis, enables our approach to scale and be more efficient as the
number and complexity of graph operators increases.

\setlength\belowcaptionskip{-6pt}
\begin{figure*}
    \begin{subfigure}{.33\linewidth}
        \includegraphics[width=.9\linewidth,height=3.5cm]{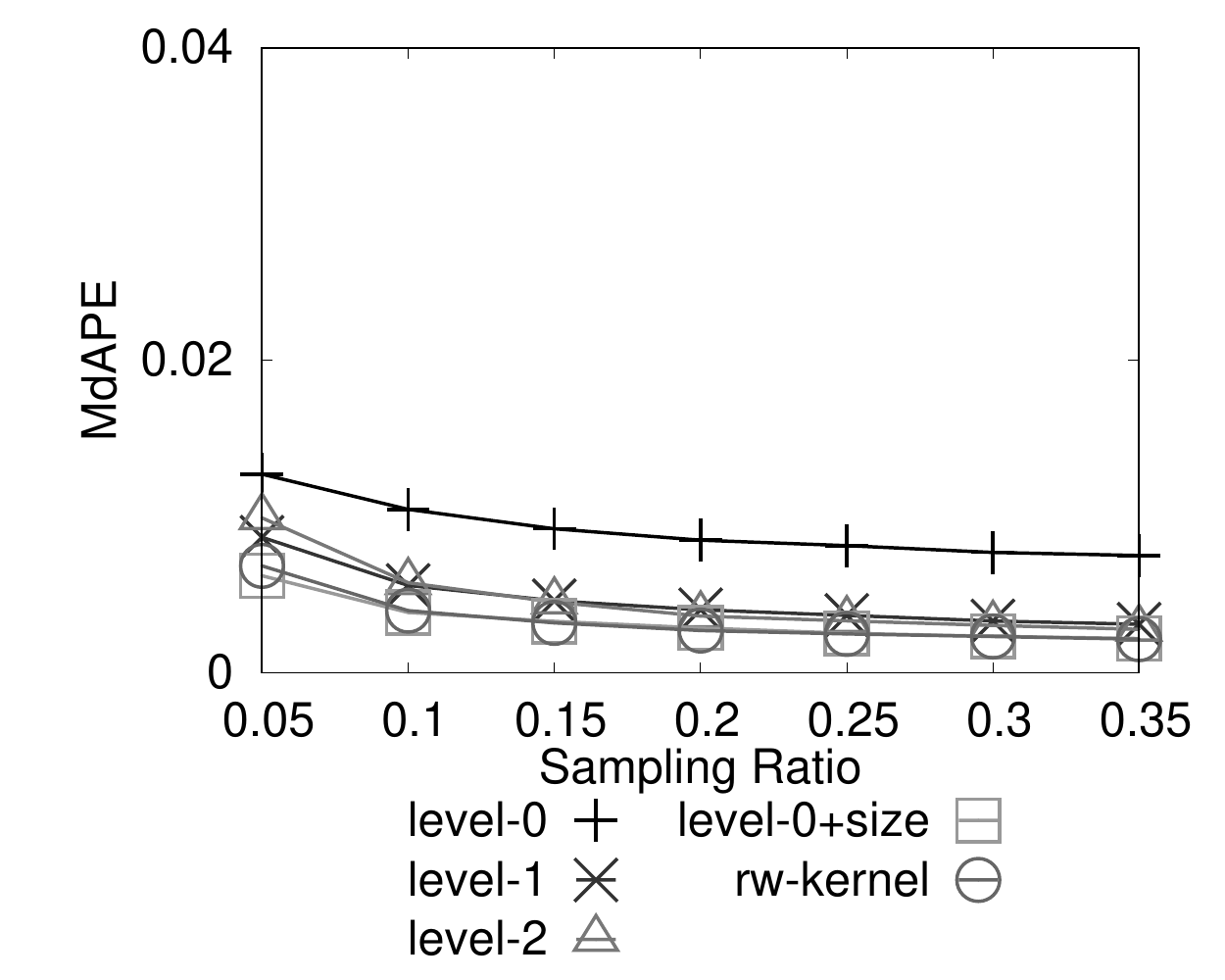}
        \vspace{-6pt}
        \caption{Spectral Radius}
        \label{figure:sme_as_er_mdape}
    \end{subfigure}%
    \begin{subfigure}{.33\linewidth}
        \includegraphics[width=.9\linewidth,height=3.5cm]{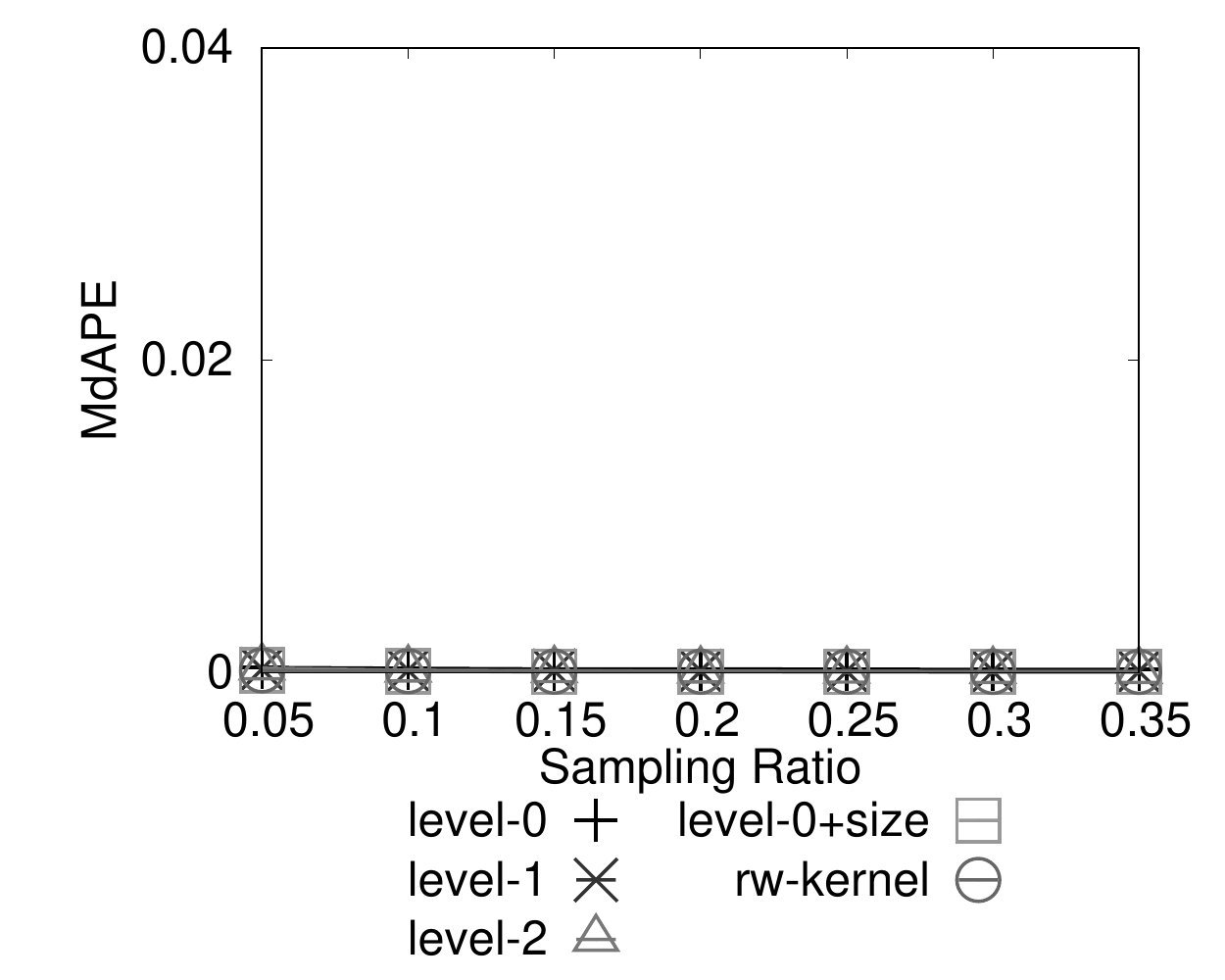}
        \vspace{-6pt}
        \caption{Eigenvector C.}
        \label{figure:sme_as_ec_mdape}
    \end{subfigure}%
    \begin{subfigure}{.33\linewidth}
        \includegraphics[width=.9\linewidth,height=3.5cm]{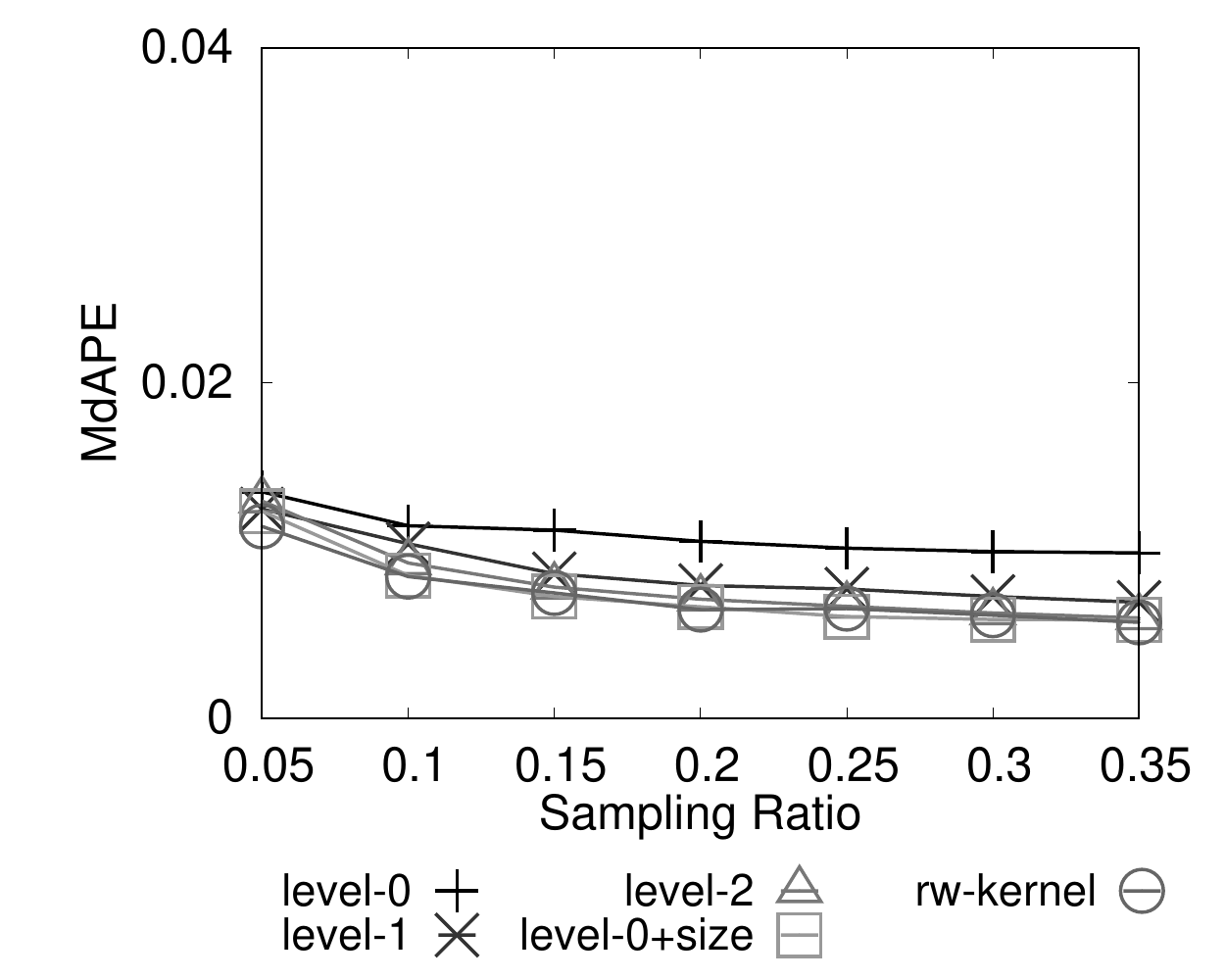}
        \vspace{-6pt}
        \caption{Betweenness C.}
        \label{figure:sme_as_bc_mdape}
    \end{subfigure}
    \begin{subfigure}{.33\linewidth}
        \includegraphics[width=.9\linewidth,height=3.5cm]{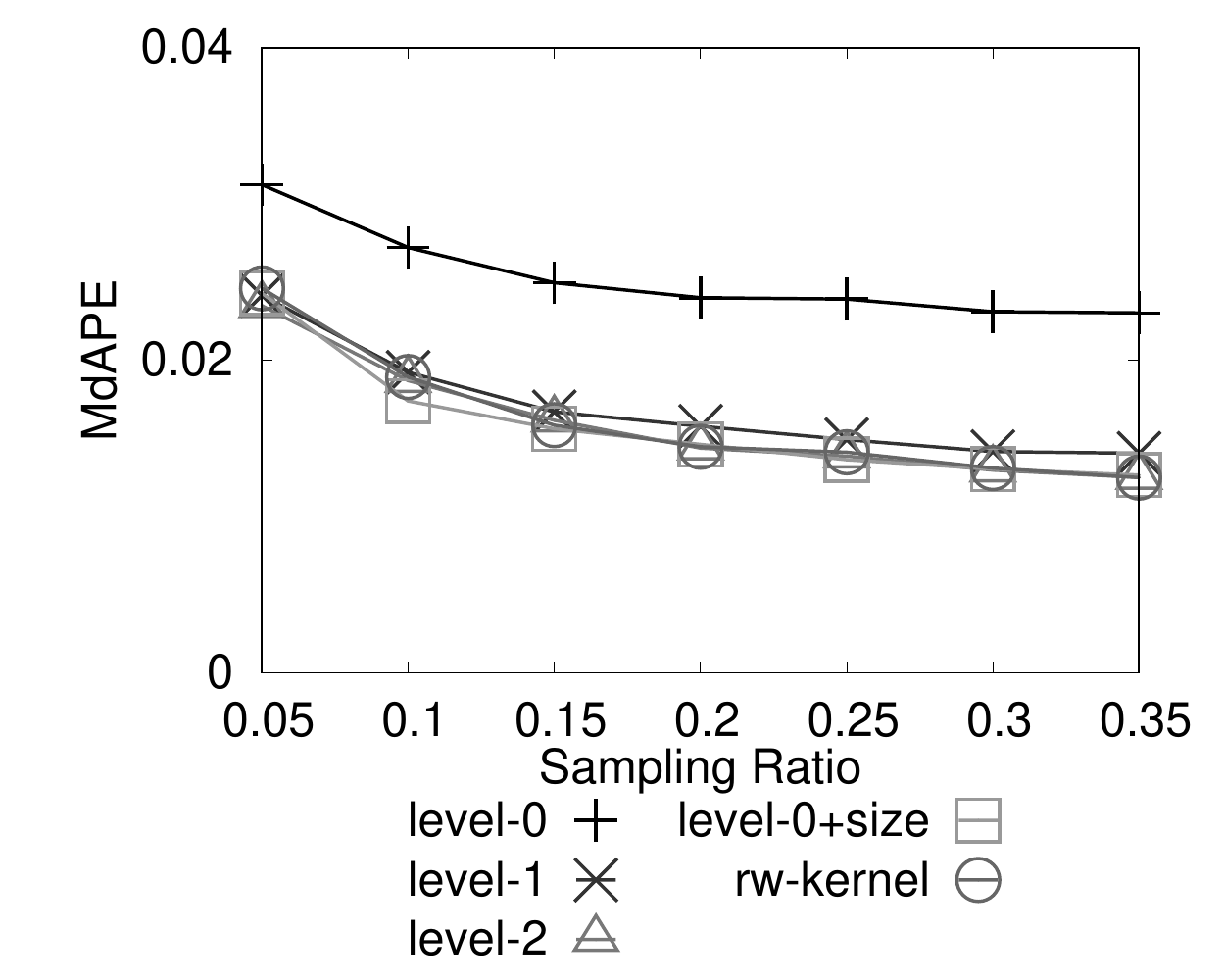}
        \vspace{-6pt}
        \caption{Edge Betweenness C.}
        \label{figure:sme_as_ebc_mdape}
    \end{subfigure}%
    \begin{subfigure}{.33\linewidth}
        \includegraphics[width=.9\linewidth,height=3.5cm]{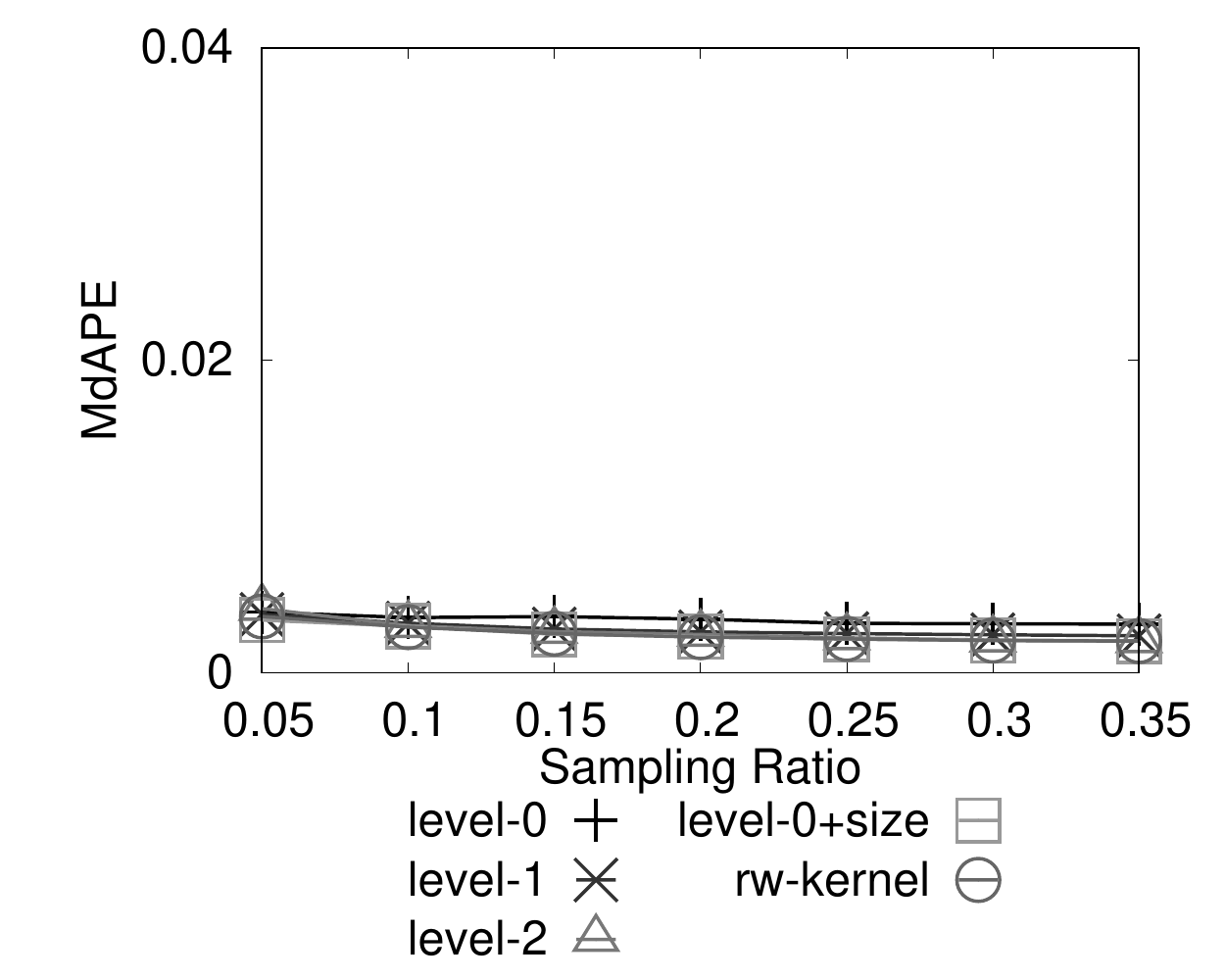}
        \vspace{-6pt}
        \caption{Closeness C.}
        \label{figure:sme_as_cc_mdape}
    \end{subfigure}%
    \begin{subfigure}{.33\linewidth}
        \includegraphics[width=.9\linewidth,height=3.5cm]{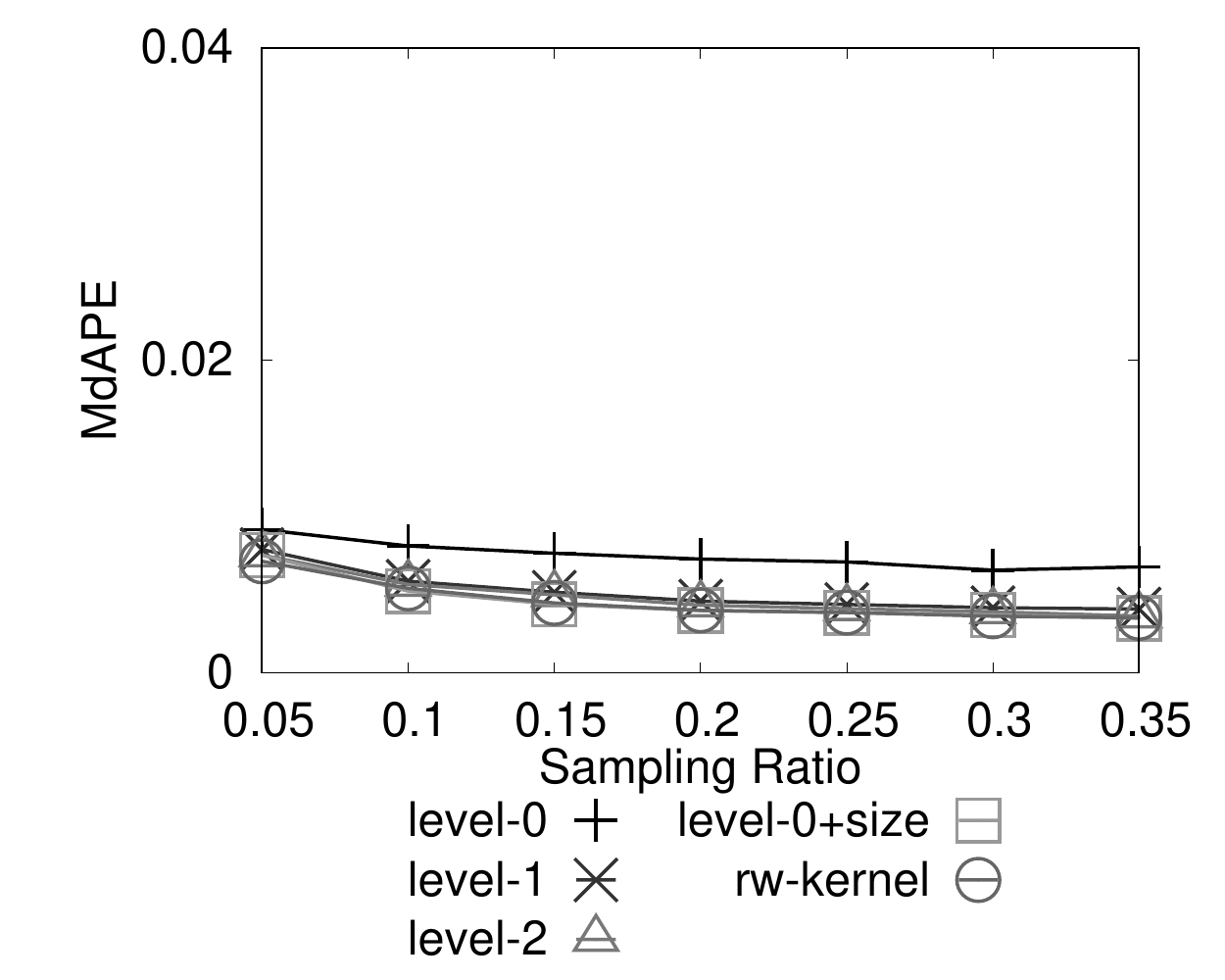}
        \vspace{-6pt}
        \caption{PageRank}
        \label{figure:sme_as_pr_mdape}
    \end{subfigure}
    \caption{Similarity Metrics Comparison (\emph{AS})}\label{figure:sme_as}
    \begin{subfigure}{.33\linewidth}
        \includegraphics[width=.9\linewidth,height=3.5cm]{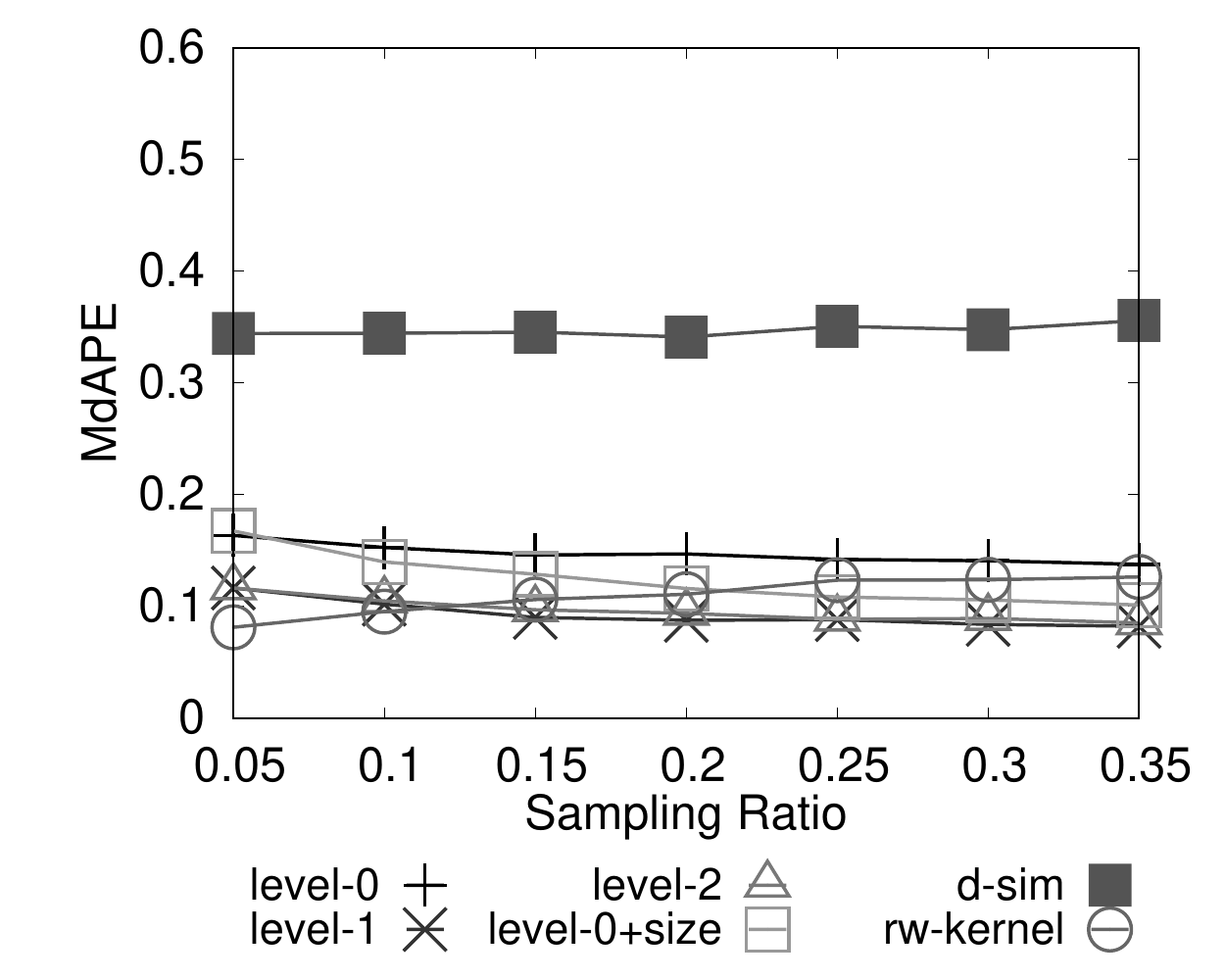}
        \vspace{-6pt}
        \caption{Spectral Radius}
        \label{figure:sme_tw_er_mdape}
    \end{subfigure}%
    \begin{subfigure}{.33\linewidth}
        \includegraphics[width=.9\linewidth,height=3.5cm]{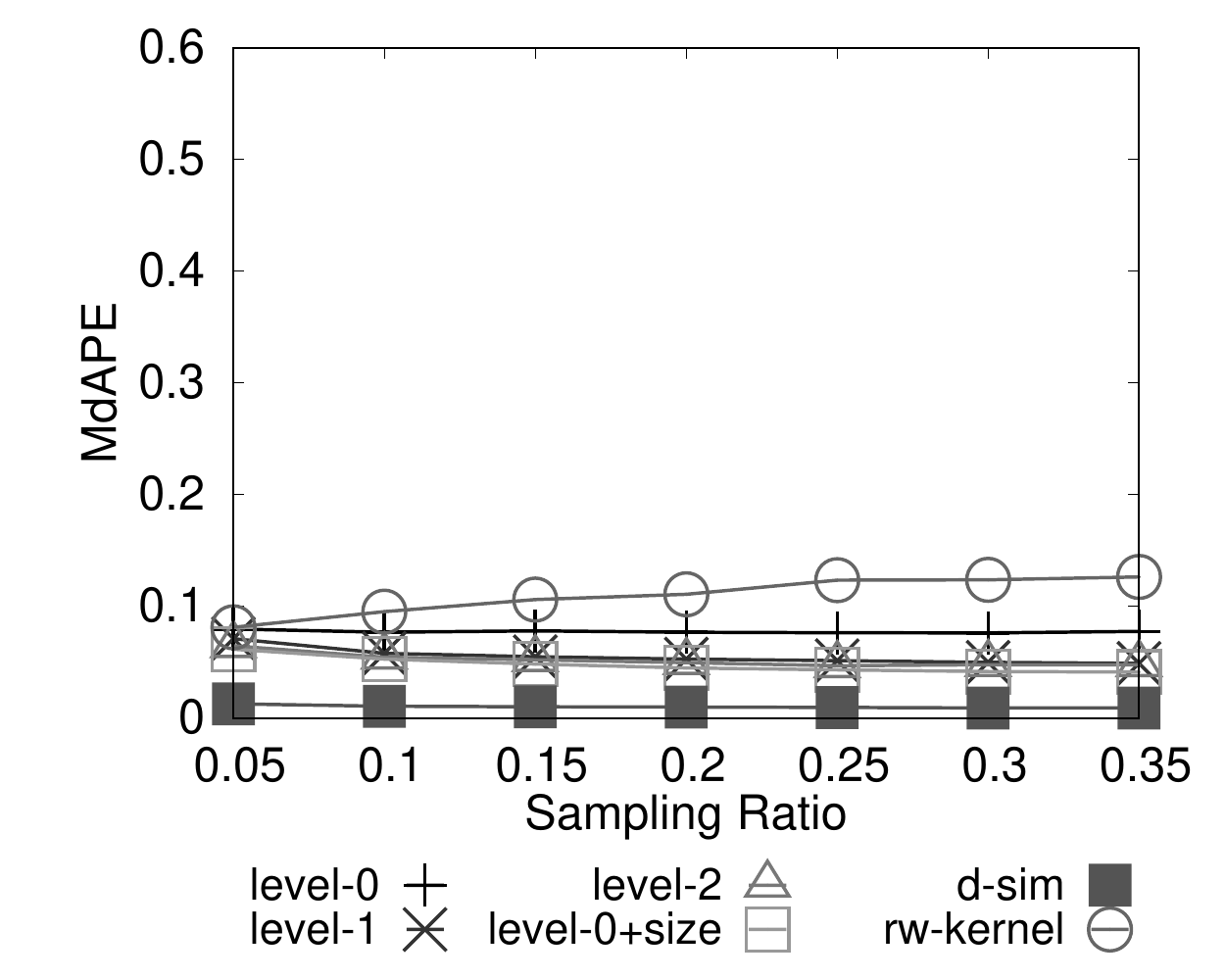}
        \vspace{-6pt}
        \caption{Eigenvector C.}
        \label{figure:sme_tw_ec_mdape}
    \end{subfigure}%
    \begin{subfigure}{.33\linewidth}
        \includegraphics[width=.9\linewidth,height=3.5cm]{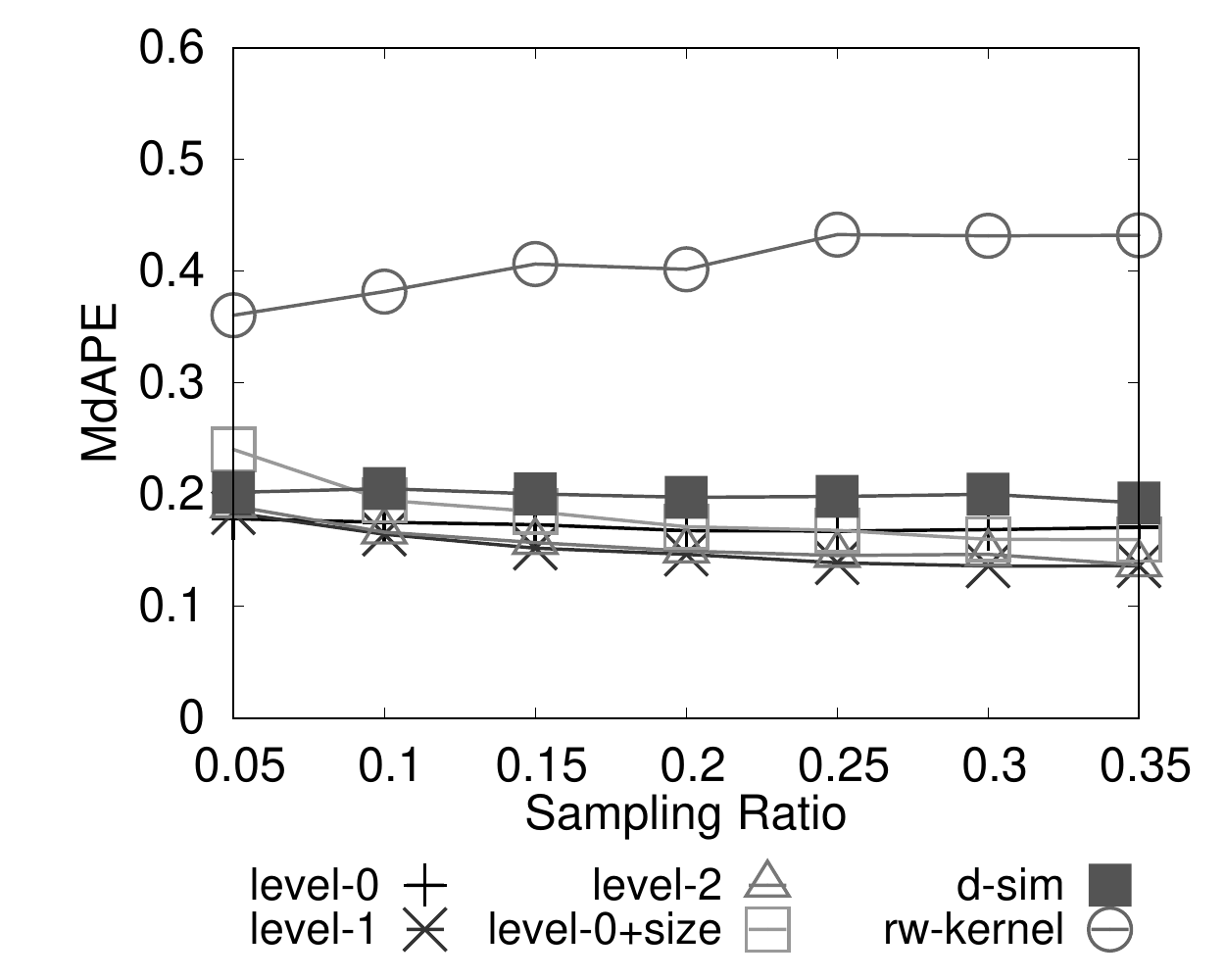}
        \vspace{-6pt}
        \caption{Betweenness C.}
        \label{figure:sme_tw_bc_mdape}
    \end{subfigure}
    \begin{subfigure}{.33\linewidth}
        \includegraphics[width=.9\linewidth,height=3.5cm]{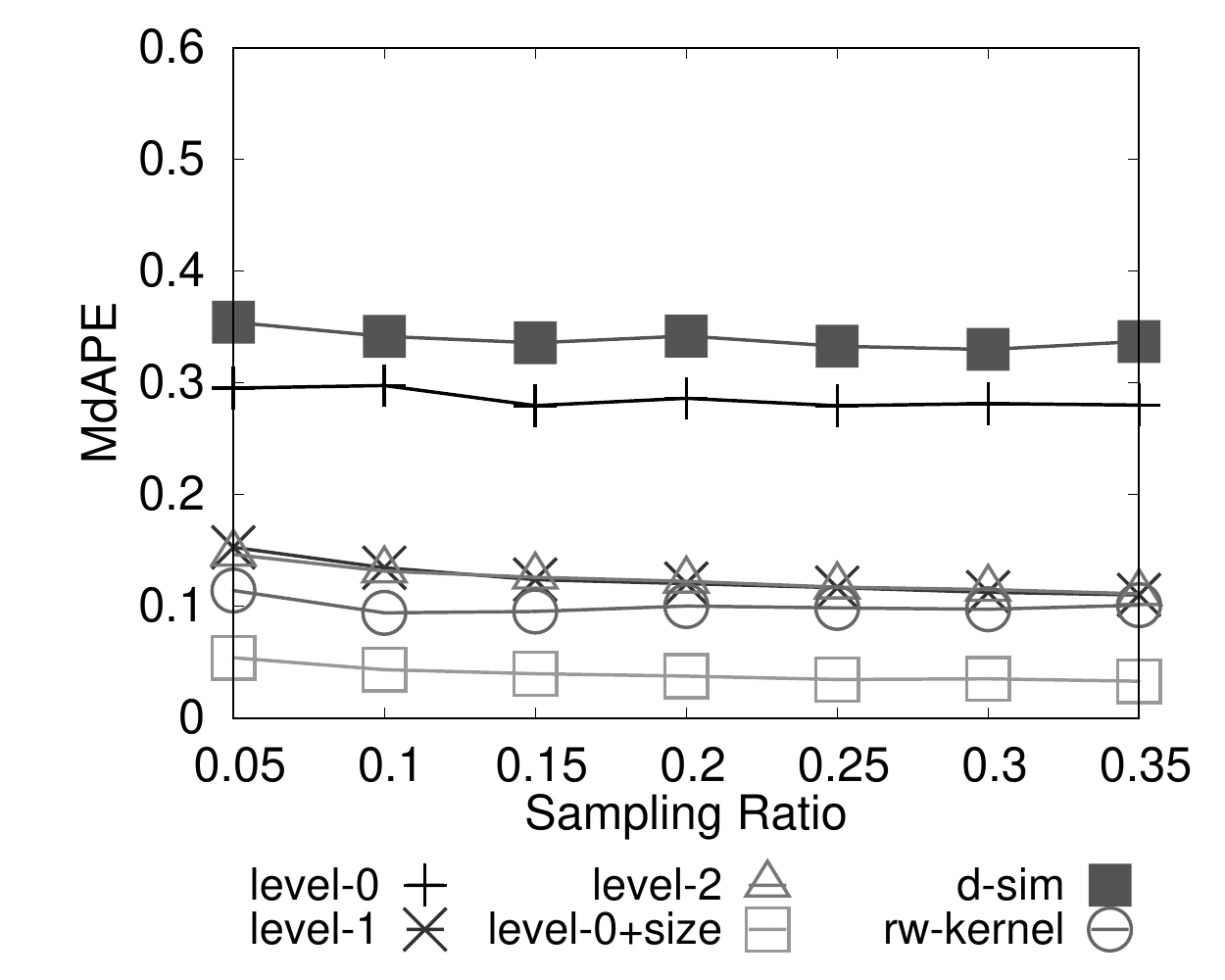}
        \vspace{-6pt}
        \caption{Edge Betweenness C.}
        \label{figure:sme_tw_ebc_mdape}
    \end{subfigure}%
    \begin{subfigure}{.33\linewidth}
        \includegraphics[width=.9\linewidth,height=3.5cm]{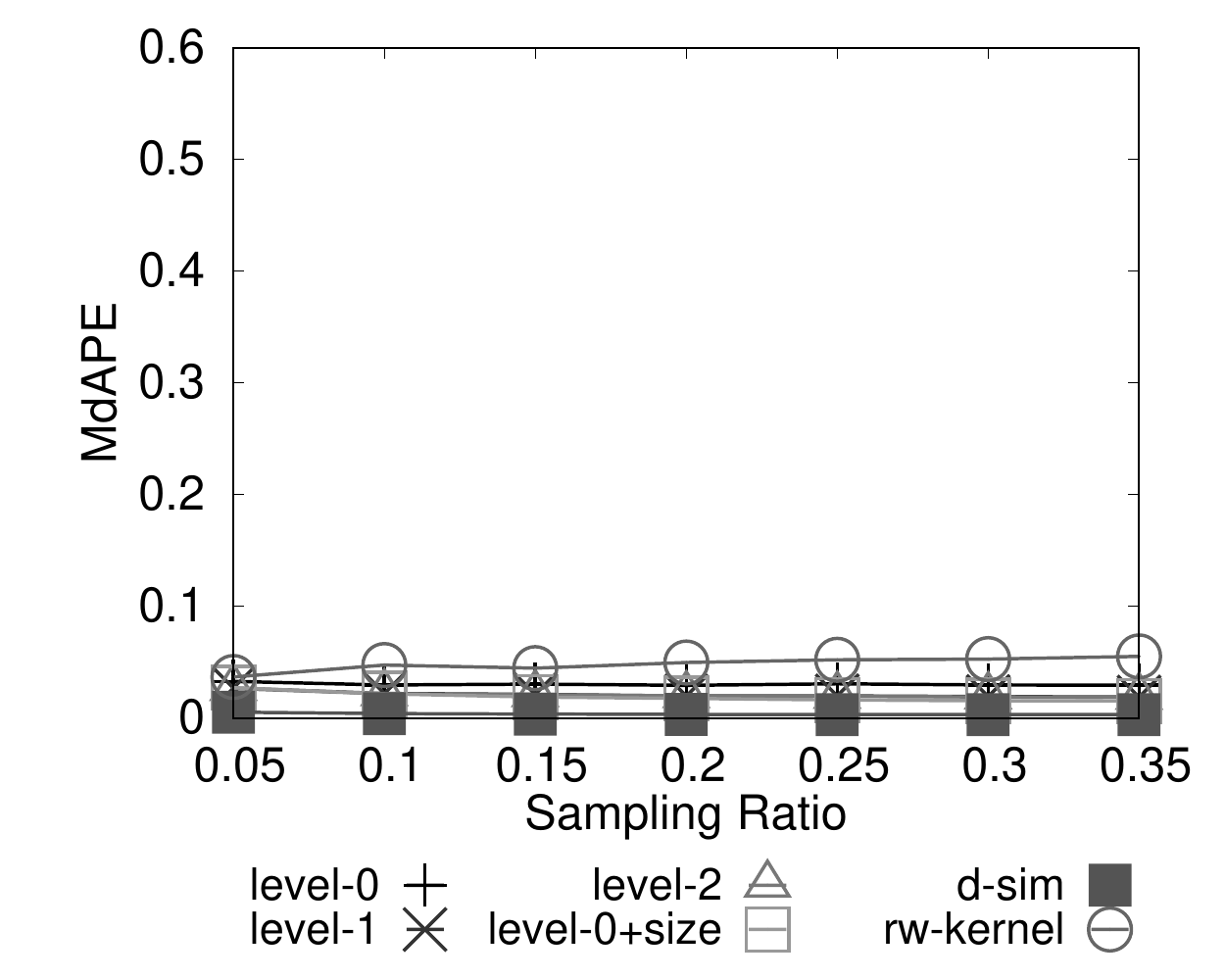}
        \vspace{-6pt}
        \caption{Closeness C.}
        \label{figure:sme_tw_cc_mdape}
    \end{subfigure}%
    \begin{subfigure}{.33\linewidth}
        \includegraphics[width=.9\linewidth,height=3.5cm]{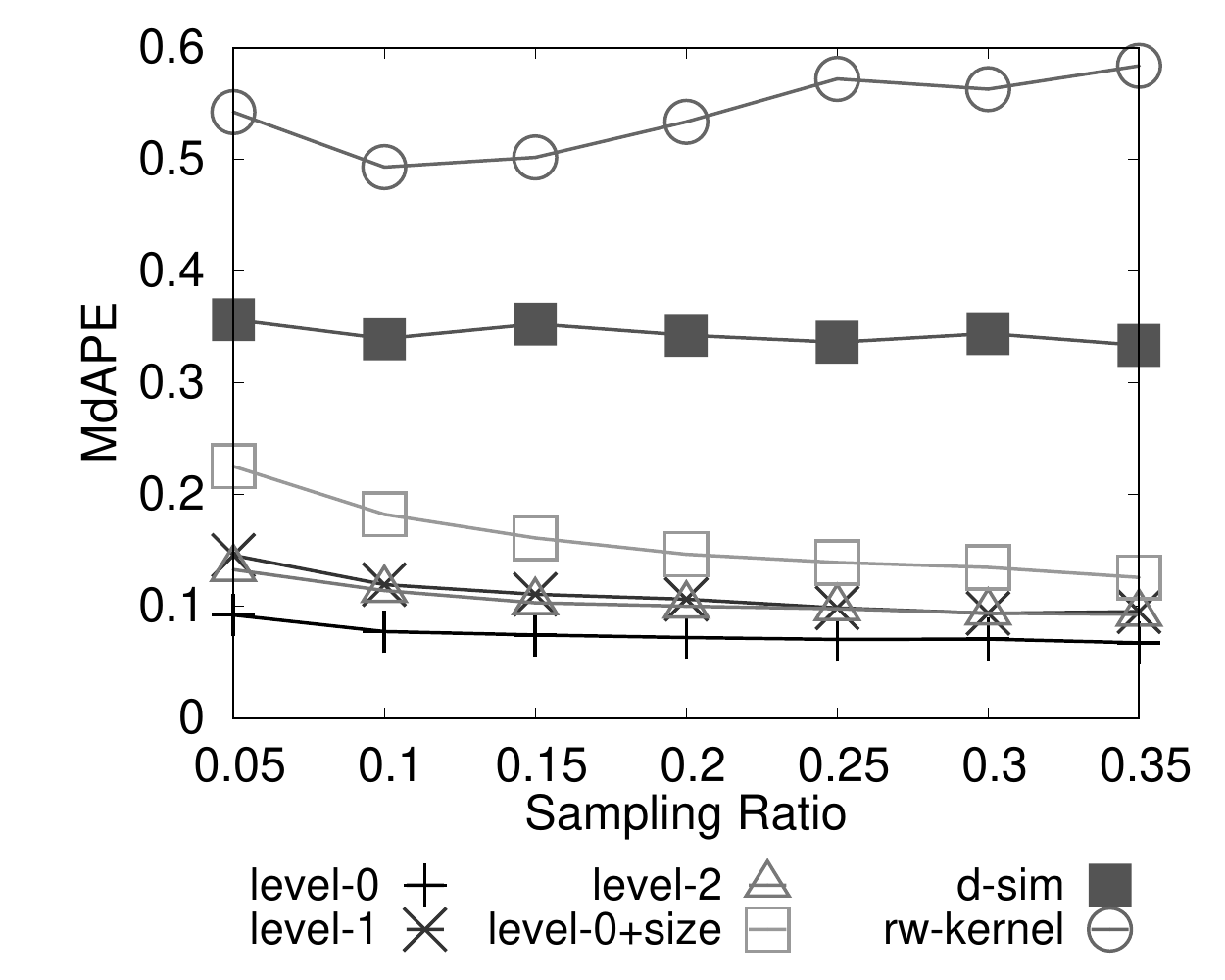}
        \vspace{-6pt}
        \caption{PageRank}
        \label{figure:sme_tw_pr_mdape}
    \end{subfigure}
    \caption{Similarity Metrics Comparison (\emph{TW})}\label{figure:sme_tw}
    \begin{subfigure}{.33\linewidth}
        \includegraphics[width=.9\linewidth,height=3.5cm]{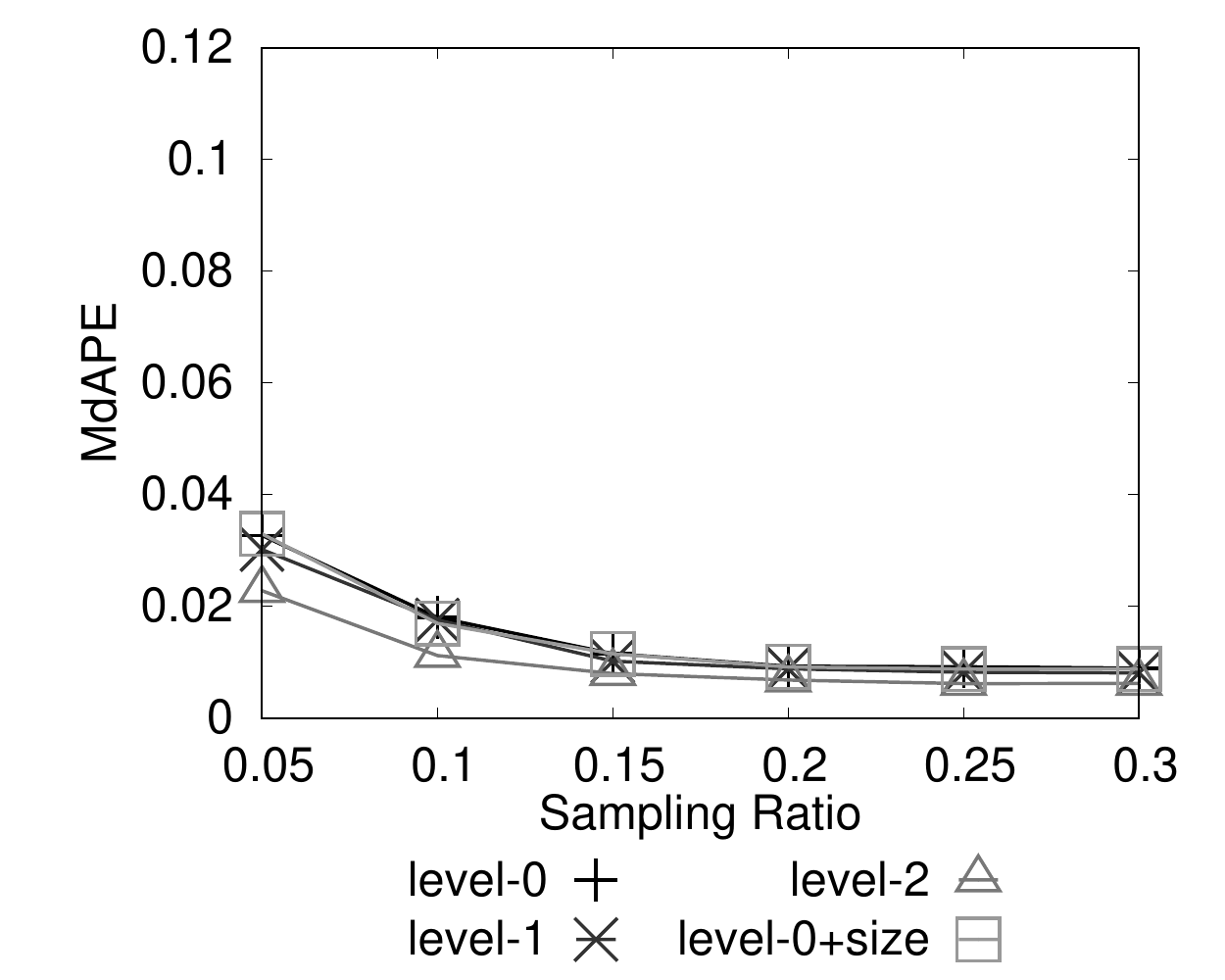}
        \vspace{-6pt}
        \caption{Spectral Radius}
        \label{figure:sme_ba_er_mdape}
    \end{subfigure}%
    \begin{subfigure}{.33\linewidth}
        \includegraphics[width=.9\linewidth,height=3.5cm]{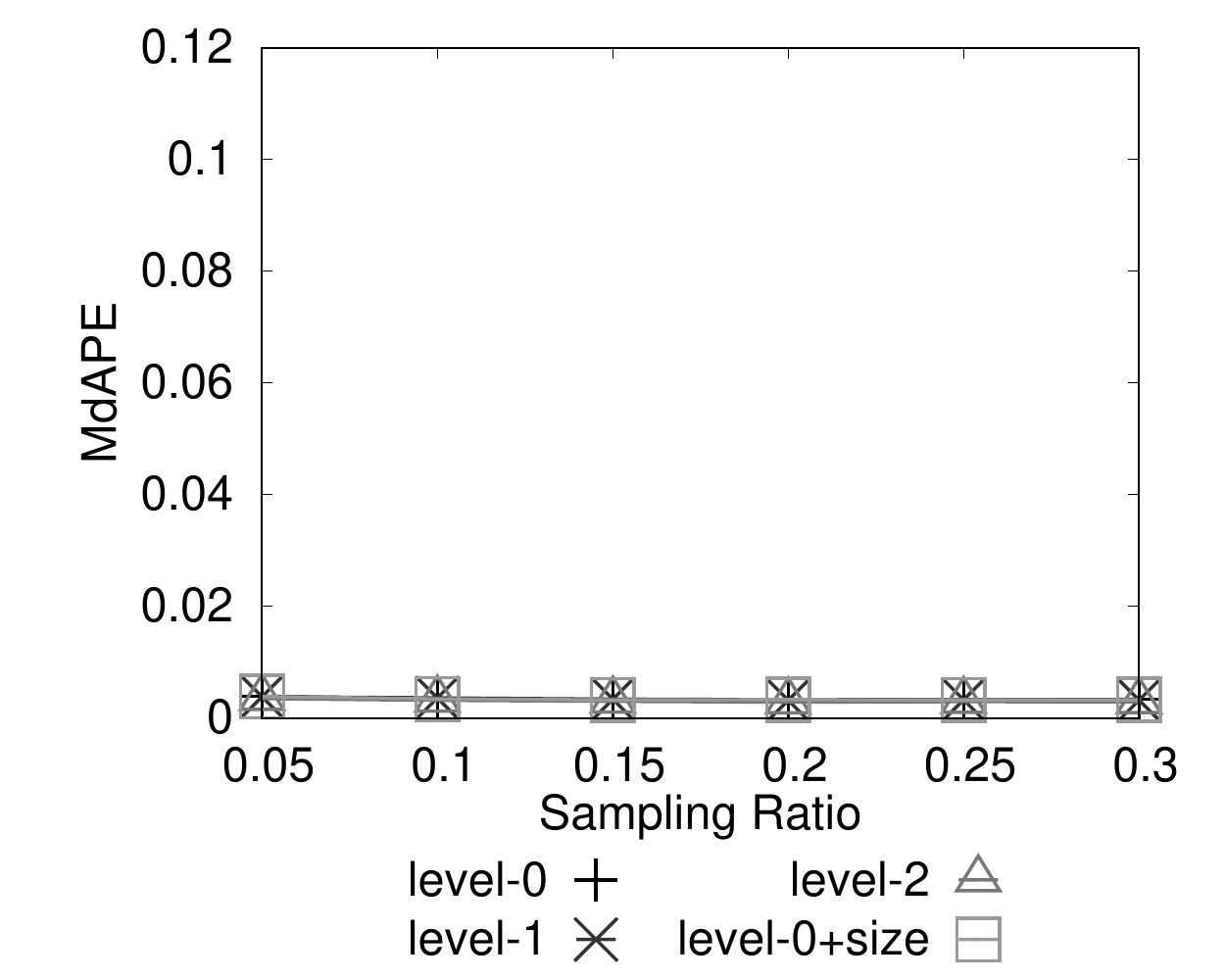}
        \vspace{-6pt}
        \caption{Eigenvector C.}
        \label{figure:sme_ba_ec_mdape}
    \end{subfigure}%
    \begin{subfigure}{.33\linewidth}
        \includegraphics[width=.9\linewidth,height=3.5cm]{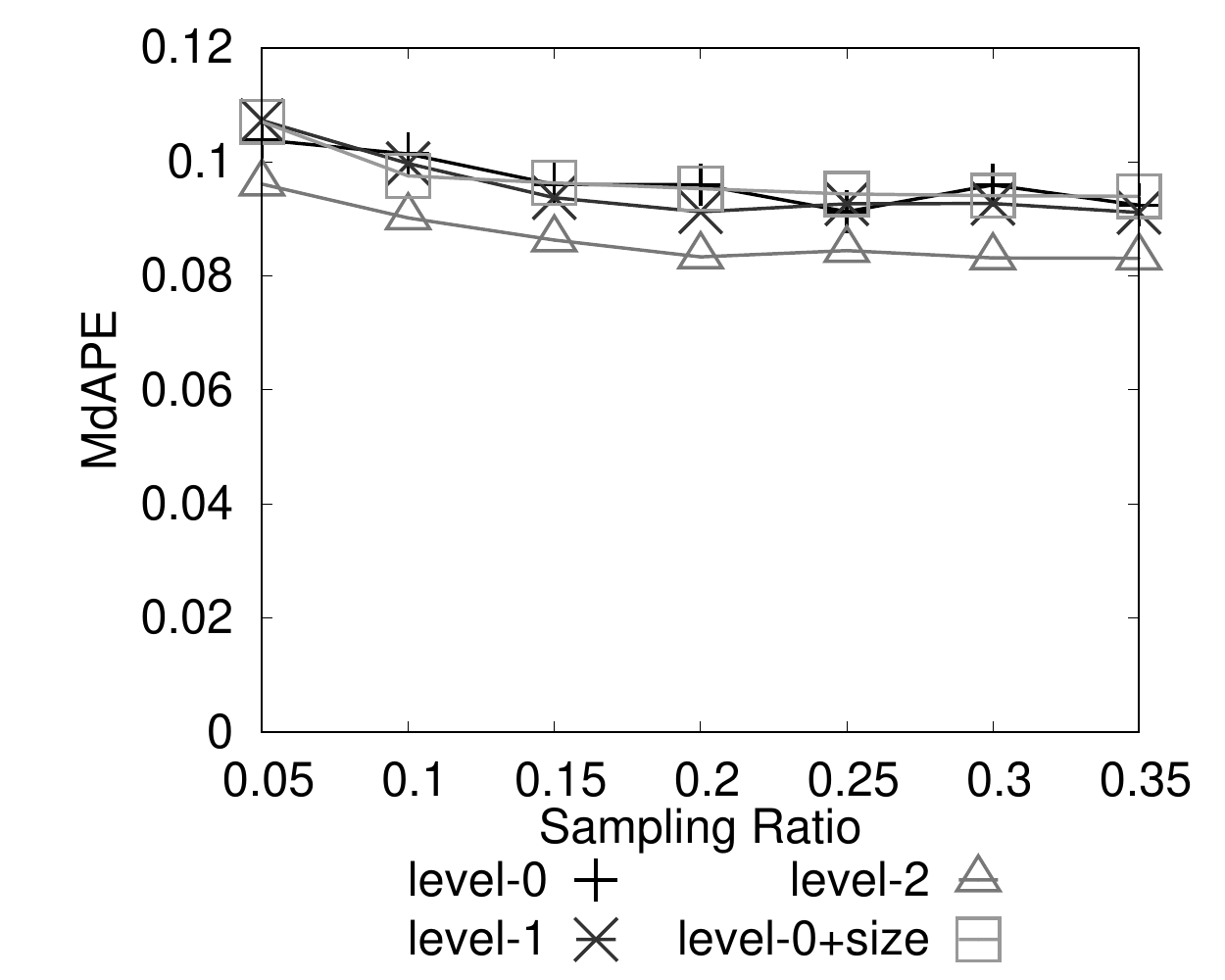}
        \vspace{-6pt}
        \caption{Betweenness C.}
        \label{figure:sme_ba_bc_mdape}
    \end{subfigure}
    \begin{subfigure}{.33\linewidth}
        \includegraphics[width=.9\linewidth,height=3.5cm]{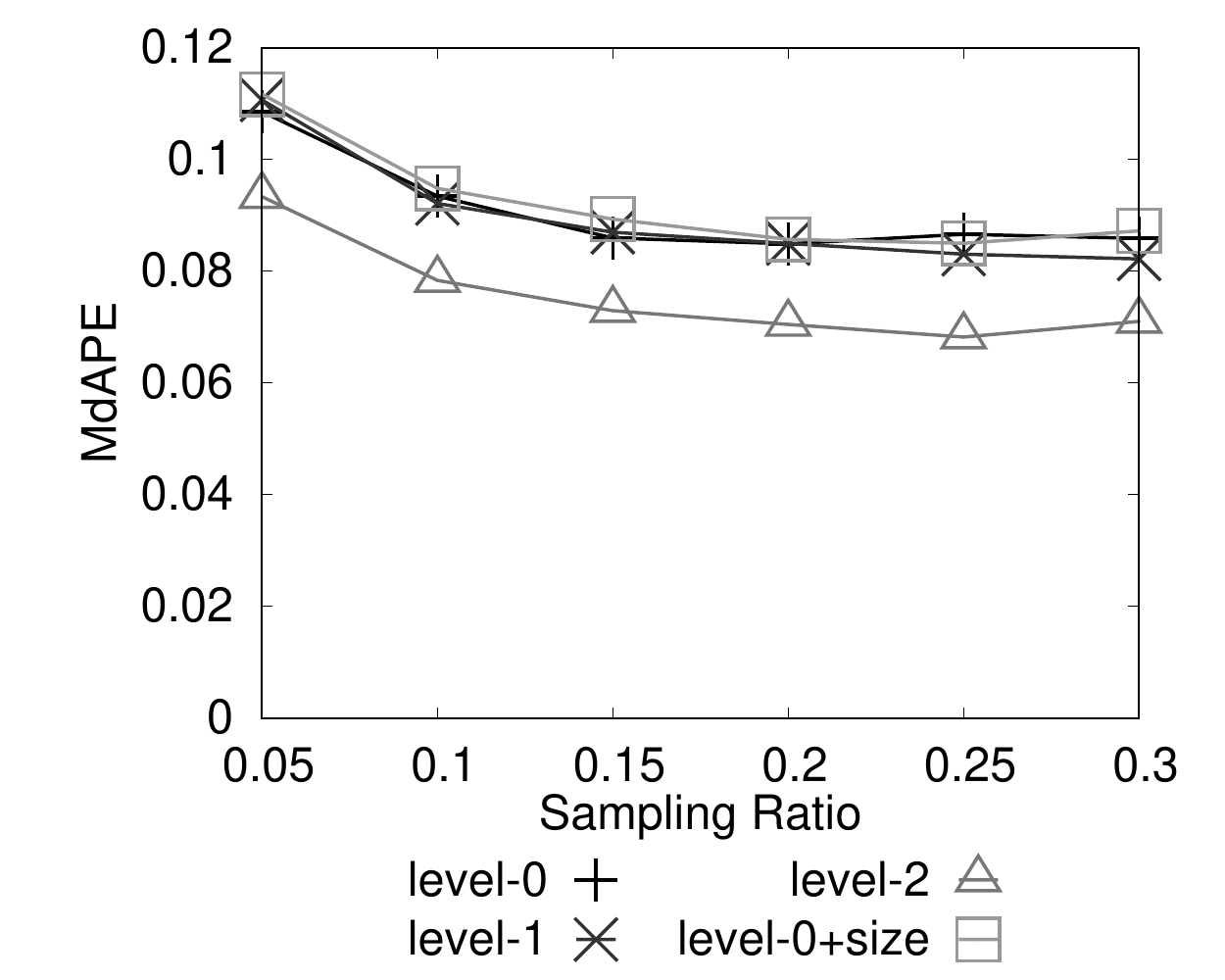}
        \vspace{-6pt}
        \caption{Edge Betweenness C.}
        \label{figure:sme_ba_ebc_mdape}
    \end{subfigure}%
    \begin{subfigure}{.33\linewidth}
        \includegraphics[width=.9\linewidth,height=3.5cm]{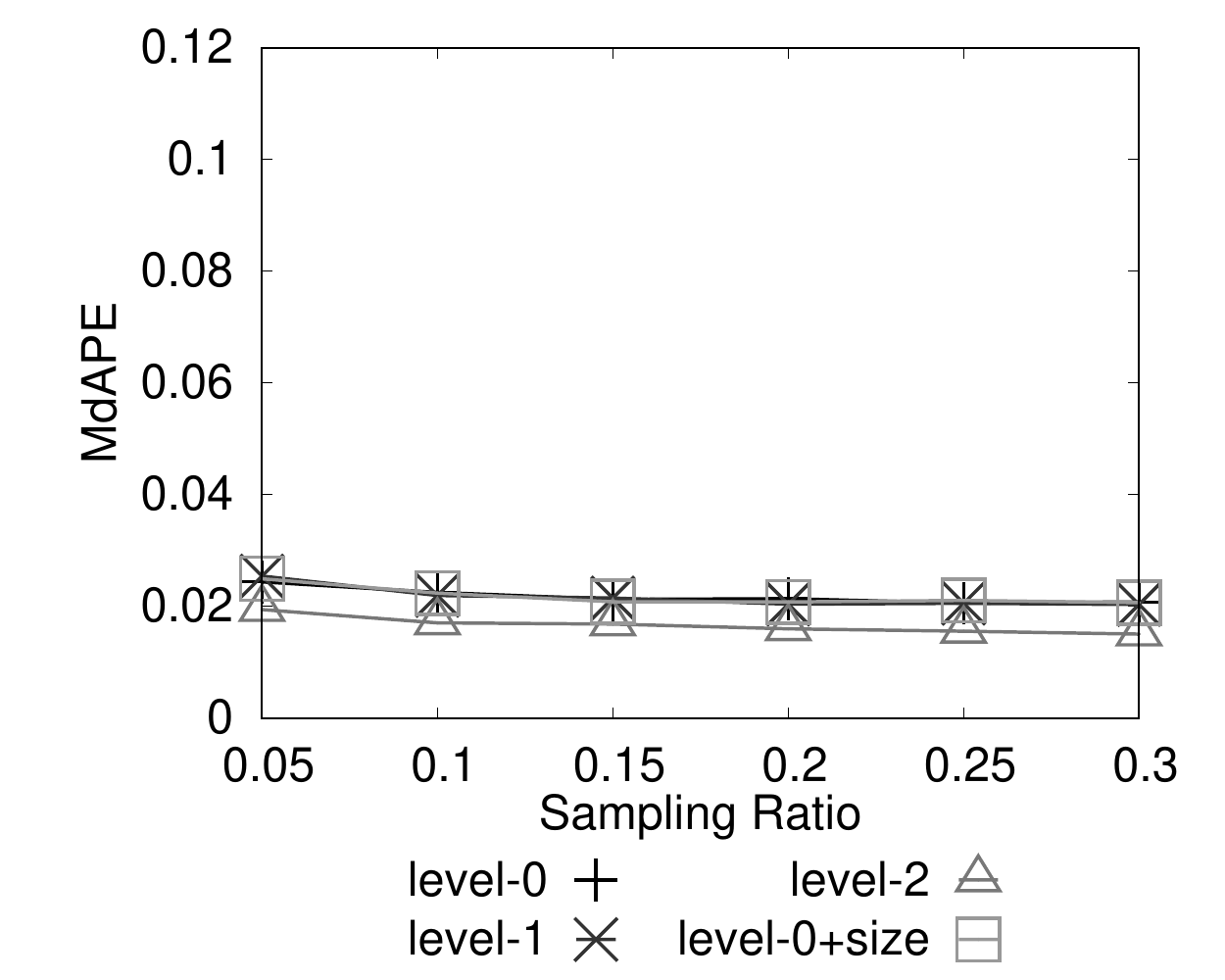}
        \vspace{-6pt}
        \caption{Closeness C.}
        \label{figure:sme_ba_cc_mdape}
    \end{subfigure}%
    \begin{subfigure}{.33\linewidth}
        \includegraphics[width=.9\linewidth,height=3.5cm]{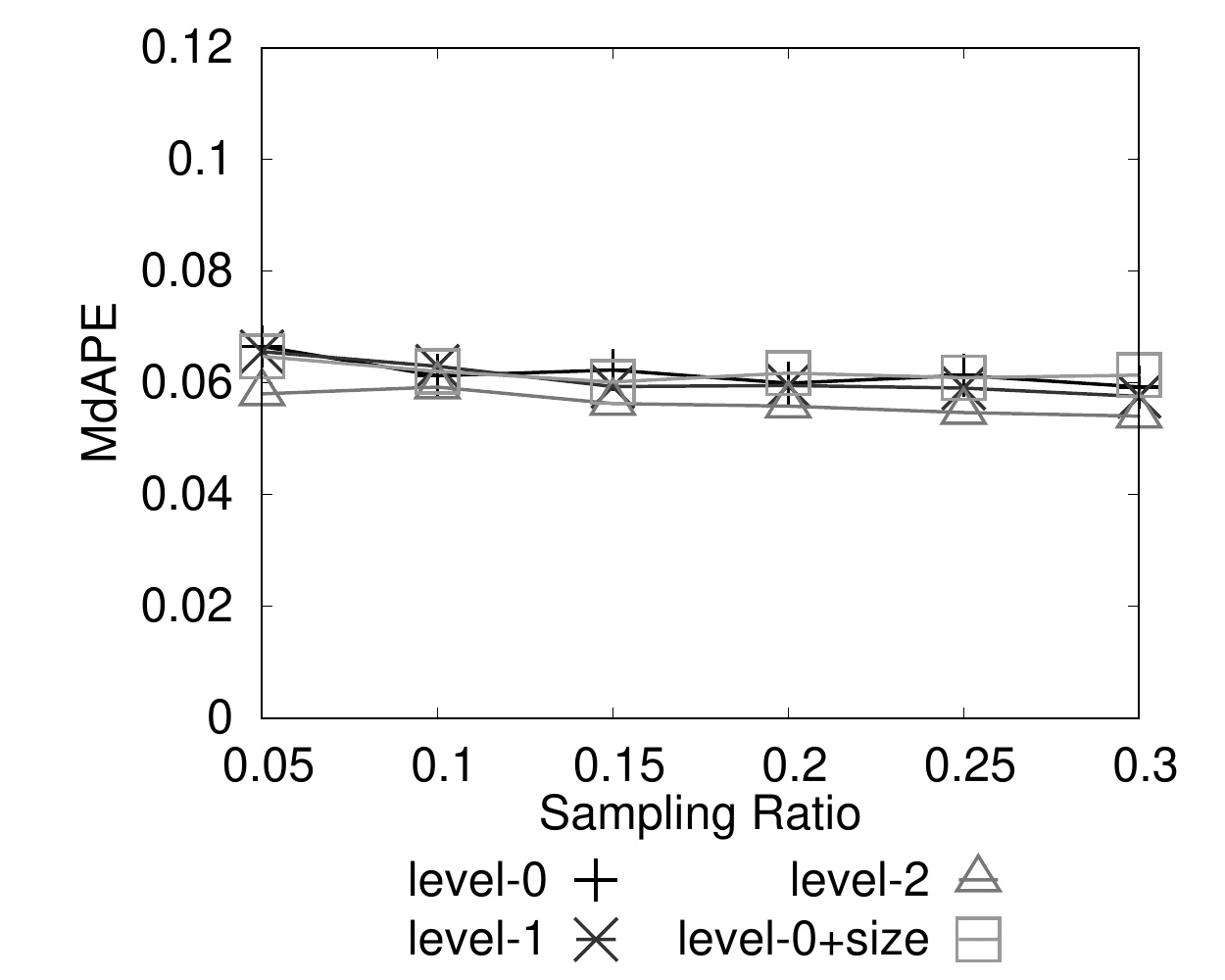}
        \vspace{-6pt}
        \caption{PageRank}
        \label{figure:sme_ba_pr_mdape}
    \end{subfigure}%
    \caption{Similarity Metrics Comparison (\emph{BA})}\label{figure:sme_ba}
\end{figure*}

\shortheader{Comparing Similarity Measures:} The results of the similarity
measure comparisons, for the \emph{AS}, \emph{TW} and \emph{BA} datasets, are
displayed in Figures \ref{figure:sme_as}, \ref{figure:sme_tw} and
\ref{figure:sme_ba}, respectively, where $MdAPE$ is used to express the
modeling error.  For the \emph{TW} dataset we compare six similarity measures:
The \emph{degree distribution + levels} measure (for levels equal from 0 to 2),
a combination of \emph{level-0} degree distribution with vertex count (denoted
by \emph{level-0 + size}), \emph{D-measure} and the \emph{Random Walk Kernel}
based similarity measure (denoted by rw-kernel). 
In the cases of \emph{AS} and 
\emph{BA} we do not include \emph{D-measure}, since it was not possible to 
compute it, for the graphs of those datasets, because of its running time. In
addition for \emph{BA} we do not include \emph{rw-kernel} for the same reason.
The results indicate the impact that the choice of
similarity measure has on modeling accuracy. A more suitable to the modeled
operator and detailed similarity measure is more sensitive to topology
differences and can lead to better operator modeling.

Focusing on the \emph{TW} dataset, we observe that in all figures, with the
exception of PageRank, the \emph{degree distribution + levels}
similarity measure, for a number of levels, can model an operator more
accurately than the simple degree distribution-based, effectively reducing the
errors reported in Table \ref{table:sampling_rate_vs_errors}. Indeed, the
addition of more levels to the degree distribution incorporates more
information about the connectivity of each vertex. This additional topological
insights contribute positively to better estimate the similarity of two graphs.
For instance, this allows the MdAPE error to drop from 29.5\% to about 15\%
when utilizing a \emph{level-2} similarity for edge betweenness centrality and
$p=5\%$. 

Examining the modeling quality, we observe that it increases but only up to a
certain point, in relation to the topology of the graphs in the dataset.  For
example, since \emph{TW} comprises of ego graphs, all the degrees of level $>2$
are zero, since there exist no vertices with distance greater than $2$;
therefore, employing more levels does not contribute any additional information
about the topology of the graphs when computing their similarity.

Finally, we observe that, in specific cases, such as PageRank (Figure
\ref{figure:sme_tw_pr_mdape}), enhancing the degree distribution with degrees
of more levels introduces information that is interpreted as noise during
modeling. PageRank is better modeled with the simple degree distribution as a
similarity measure. As such, we argue that for a given dataset and graph
operator, experimentation is required to find the number of levels that
give the best tradeoff between accuracy and execution time.

We next concentrate on the effect of the combination of degree distribution
with vertex count in the modeling accuracy. We note that the vertex count
contributes positively in the modeling of distance-related metrics while having
a neutral or negative impact on degree- and spectrum-related metrics. This is
attributed to the existence of, at least, a mild correlation, between vertex
count and \emph{bc}, \emph{ebc} and \emph{cc}
\cite{DBLP:journals/nhm/JamakovicU08}. For our least accurately approximated
task, edge betweenness centrality, employing the combination of measures results
in a more than 6$\times$ decrease in error.

For \emph{D-measure}, our experiments show that, for distance-related metrics
it performs at least as good as the \emph{degree distribution + levels}
similarity measures for a given level, with the notable exception of the
PageRank case. On the other hand, the degree distribution can be sufficiently
accurate for degree- or spectrum-related metrics. As \textit{D-measure} is
based on distance distributions between vertices, having good accuracy for
distance-related measures is something to be expected.  However, \emph{degree
distribution + levels} measures exhibit comparable accuracy for distance-related
metrics as well. A good example of the effectiveness of \textit{D-measure} is
shown in the case of closeness centrality that involves all-pairs node distance
information directly incorporated in \textit{D-measure} as we have seen in
Section \ref{experimental_setup:similarity_measures}. In Figure
\ref{figure:sme_tw_cc_mdape} we observe that by adding levels we get better
results, vertex count contributes into even better modeling but
\textit{D-measure} gives better approximations. Yet, our methods' errors are
already very small (less than 3\%) in this case.  Considering the
\emph{rw-kernel} similarity measure, we observe that it performs poorly for
most of the modeled operators. Although its modeling accuracy is comperable to
the \emph{degree distribution + levels} similarity measures for some operators,
we find that for a certain level or in combination with vertex count a degree
distribution-based measure has better accuracy. Notably, \emph{rw-kernel} has
low accuracy for degree and distance related operators while performing
comperably in the case of spectrum operators.

Identifying betweenness centrality as one of the hardest operators to model
accurately, we note that, for the \emph{AS} and \emph{BA} datasets, the
approximation error is below $12\%$ and that the \emph{degree distribution +
levels} measures further improve on it for both datasets. Compared to \emph{TW}
(Fig. \ref{figure:sme_tw}), we observe that the \emph{level-2} similarity measure
provides better results for \emph{AS} and \emph{BA} but not \emph{TW},
something we attribute to the fact that \emph{TW} consists of ego graphs with
their \emph{level-2} degree being equal to zero.  Finally, it is expected that
\emph{level-0 + size} for \emph{BA} to be no different than plain
\emph{level-0} since all the graphs in \emph{BA} have the same vertex count by
construction.

\begin{figure}[h!]
    \includegraphics[width=.9\linewidth]{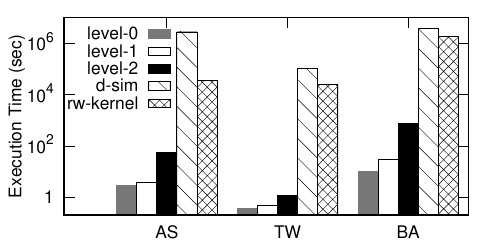}
    \caption{Execution Time (sec)}
    \label{figure:smet_plot0}
\end{figure}

Although the above similarity measures are comparable in modeling accuracy, they
are not in execution time.  A comparison in computation time for different
levels of the \emph{degree distribution + levels} similarity measure is
presented in Figure \ref{figure:smet_plot0}.  In the case of
\textit{D-measure}, the actual execution time is presented for the \emph{TW}
dataset, since it was prohibitively slow to compute it for the other two
datasets. For the remaining two datasets, we have computed \textit{D-measure}
on a random number of pairs of graphs and then projected the mean computation
time to the number of comparisons performed by our method for each dataset.

\begin{figure*}[t!]
    \begin{subfigure}{.25\linewidth}
        \centering
        \includegraphics[width=.9\linewidth]{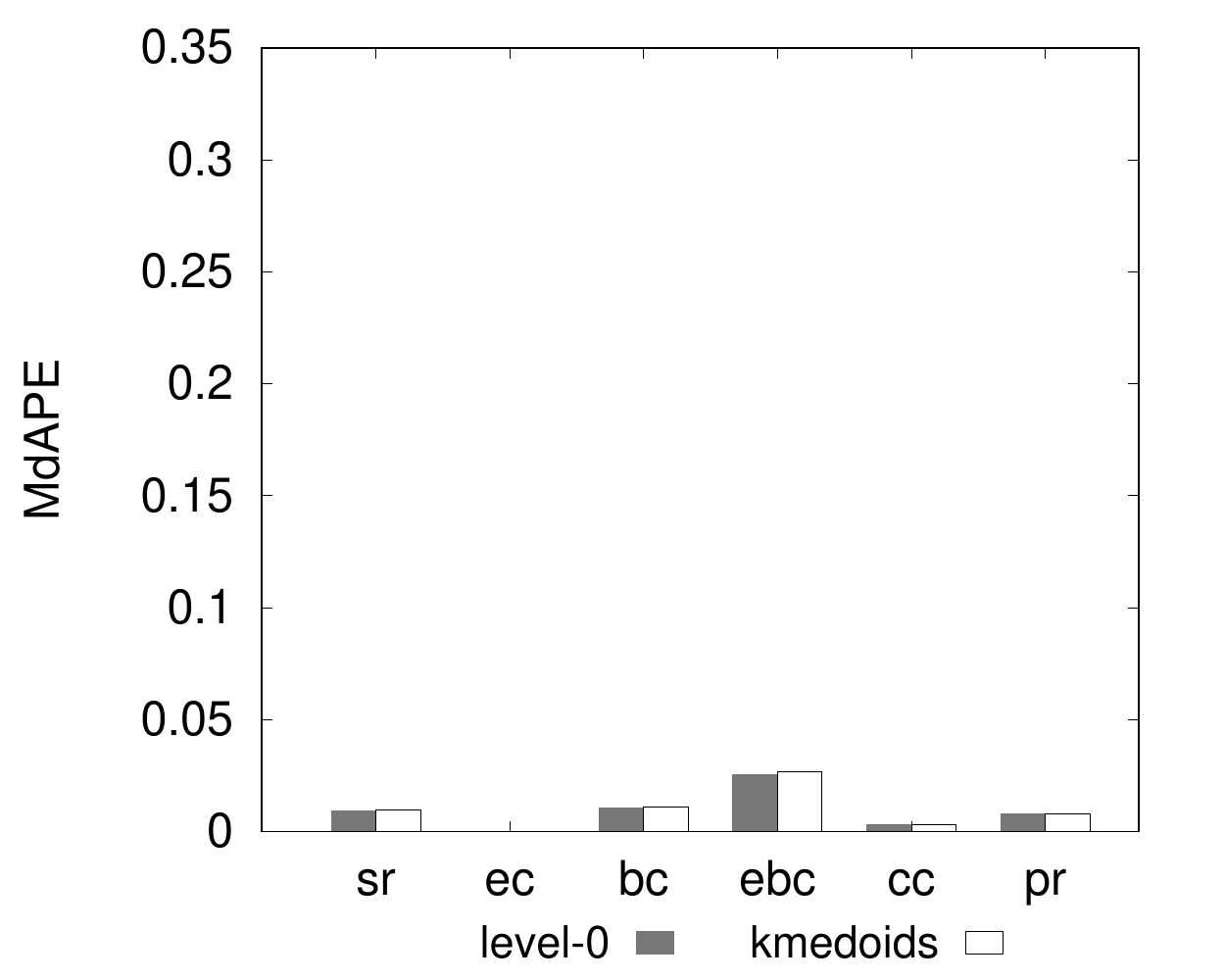}
        \caption{AS ($p = 15\%$)}
        \label{figure:aps_vs_clustering_as}
    \end{subfigure}%
    \begin{subfigure}{.25\linewidth}
        \centering
        \includegraphics[width=.9\linewidth]{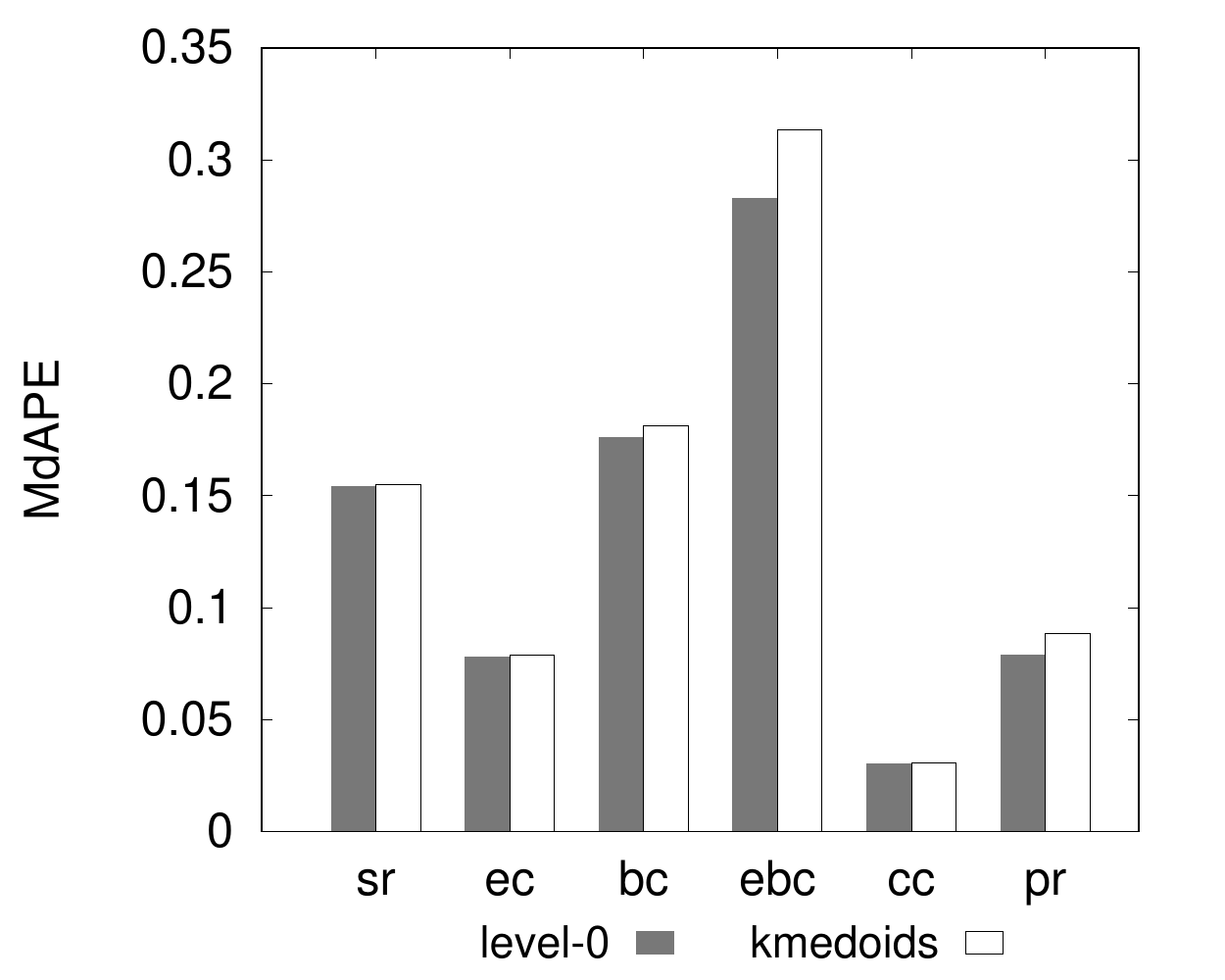}
        \caption{TW ($p = 10\%$)}
        \label{figure:aps_vs_clustering_tw}
    \end{subfigure}%
    \begin{subfigure}{.25\linewidth}
        \centering
        \includegraphics[width=.9\linewidth]{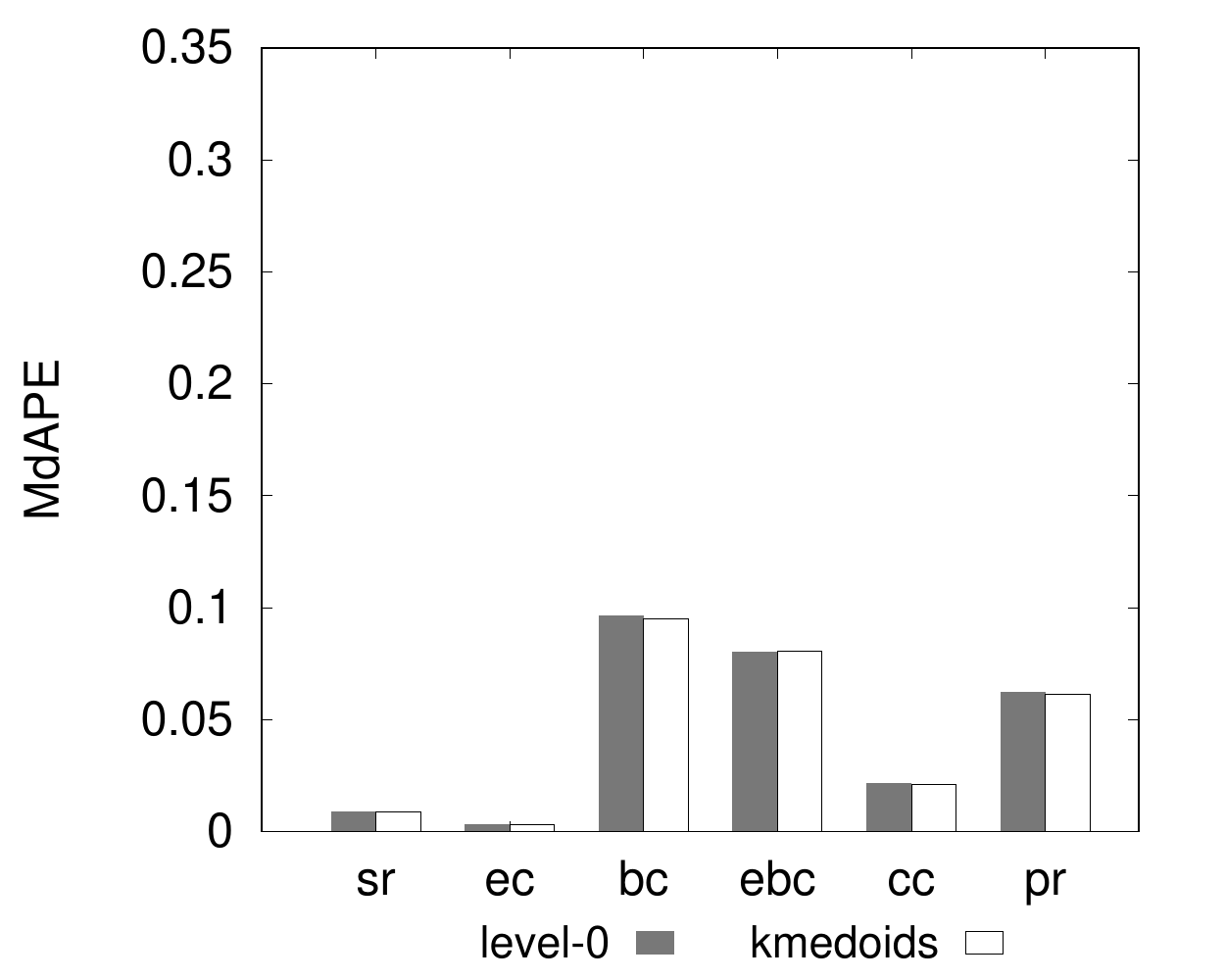}
        \caption{BA10k ($p = 3\%$)}
        \label{figure:aps_vs_clustering_ba10k}
    \end{subfigure}%
    \begin{subfigure}{.25\linewidth}
        \centering
        \includegraphics[width=.9\linewidth]{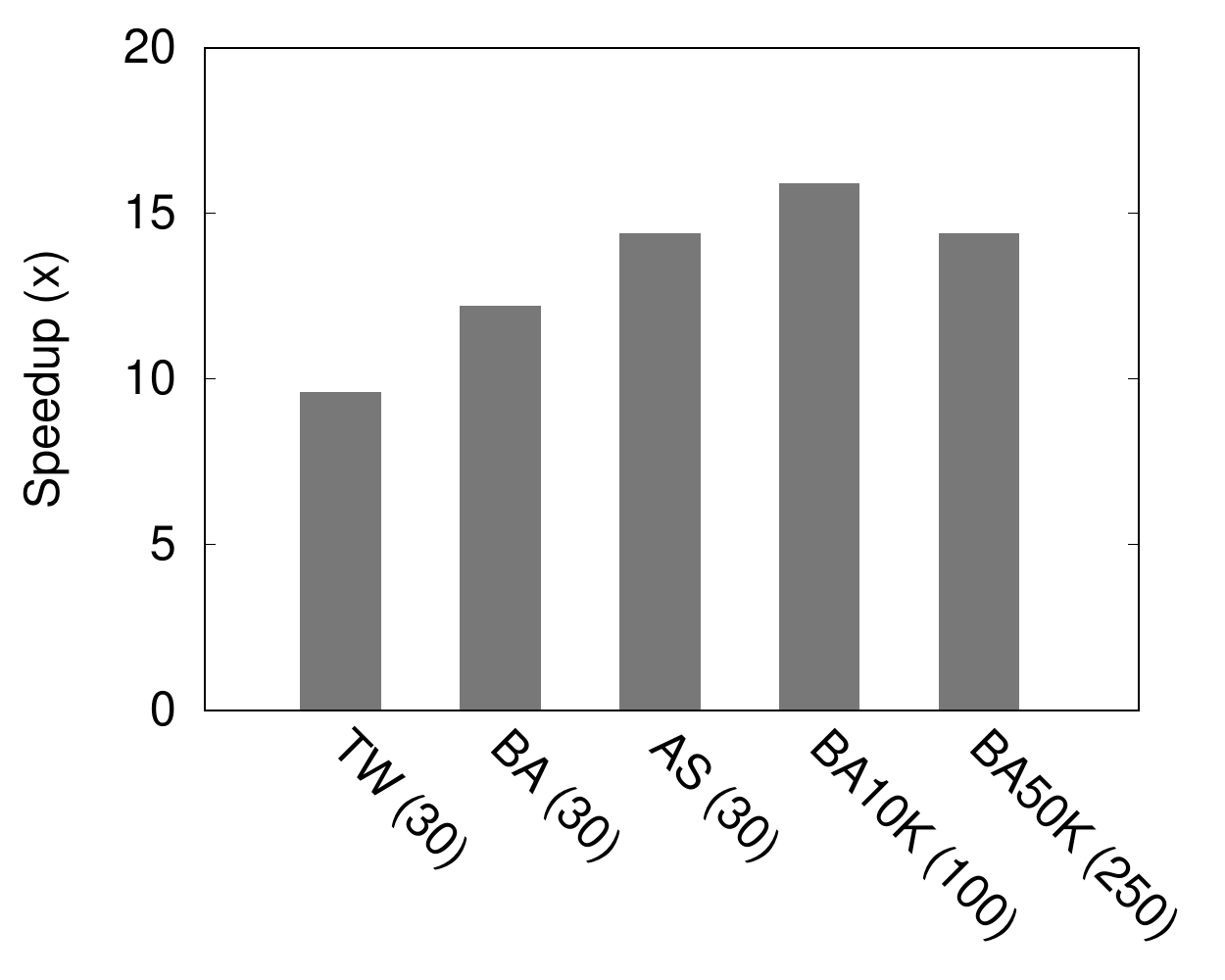}
        \caption{Similarity Computations}
        \label{figure:aps_vs_clustering_computations}
    \end{subfigure}
    \caption{All-Pairs Similarities vs Clustering}\label{figure:aps_vs_clustering}
\end{figure*}

Our results show that the overhead from \emph{level-0} to
\emph{level-1} is comparable for all the datasets. However, that is not the
case for \emph{level-2}. The higher the level, the more influential the degree
of the vertices becomes in the execution time. Specifically, while we find
\emph{level-0} to be $3.2\times$ faster than \emph{level-2} for \emph{TW}, we
observe that in the case of \emph{AS} and \emph{BA} it is $19\times$ and
$76\times$ faster.  The computation of the \textit{D-measure} and the
\textit{rw-kernel}, on the other hand, are orders of magnitude slower, i.e., we
find \emph{level-0} to be about 385K times faster than \textit{D-measure} for
the \emph{TW} dataset, while it is 273K and 933K times faster for the \emph{BA}
and \emph{AS} datasets, respectively. Given the difference in modeling quality
between the presented similarity functions, we observe a clear tradeoff between
quality of results and execution time in the context of our method.

\shortheader{Similarity Matrix Computation Speedup:} In order to evaluate the
effectiveness of the similarity matrix optimization we outlined in section
\ref{methodology:scalability}, in Fig. \ref{figure:aps_vs_clustering} we
present experimental results both in terms of modeling accuracy and speedup.
As we aim at evaluating the scalability of our method we introduce two
larger synthetic datasets \emph{BA10k} and \emph{BA50k}. They are both
created with the same settings used for the \emph{BA} dataset and contain 10k and 50k graphs respectively.  

In Fig. \ref{figure:aps_vs_clustering_as}, \ref{figure:aps_vs_clustering_tw}
and \ref{figure:aps_vs_clustering_ba10k} we compare the modeling error between
all-pairs similarity matrix (\emph{level-0}) and clustered similarity matrix
(\emph{kmedoids}) for the \emph{AS}, \emph{TW} and \emph{BA10k} datasets.  As
we mentioned in Section \ref{experiments:modeling_accuracy}, we calculate
\textit{MdAPE} and \textit{nRMSE} for a randomized $20\%$ of our datasets.  For
each dataset we use a sampling ratio $p$ based on the parameter $k = 3$ of kNN and
the number of clusters we intend to create which we set to $\sqrt{N}$. Thus,
for a dataset of 10k graphs we create 100 clusters and require $300$ graph
samples, therefore we set the sampling ratio to $3\%$. The graph samples are
randomly chosen ensuring that we have at least $k$ from each cluster. In figure
\ref{figure:aps_vs_clustering_computations} we present speedups in the number
of similarity computations performed in each of the above cases for all
datasets. We observe that the clustering optimization
adds at most a $3\%$ error in the case of \emph{TW} while achieving a
$10\times$ speedup when we use $30$ clusters. Focusing on the \emph{BA10k} and
\emph{BA50k} datasets, we observe marginal error increases and speedups up to $15\times$.
Consequently, we argue that this optimization allows our method to scale
efficiently to large graph datasets.

\section{Related Work}\label{related_work}

Our work relates to the actively researched areas of graph similarity, graph
analytics and machine learning.

\subsection{Graph Similarity}\label{related_work:graph_similarity}

The problem of determining the degree of similarity between two graphs (or
networks) has been well studied. The available techniques for quantifying graph
similarity can be classified into three main categories
(\cite{danaikoutra2011algorithms, DBLP:journals/appml/ZagerV08}):\\
\shortheader{Graph Isomorphism - Edit Distance:} Two graphs are considered
similar if they are isomorphic, i.e., there is a bijection between the vertex
sets of the two graphs such that two vertices of one graph are adjacent if and
only if their images are also adjacent \cite{bondy2011graph}. A more relaxed
approach is to consider the problem of subgraph isomorphism where one graph is
isomorphic to a subgraph of the other.  A generalization of the graph
isomorphism problem is expressed through the \textit{Edit Distance}, i.e., the
number of operations, such as additions or removals of edges or nodes, that
have to be performed in order to transform one graph to the other
\cite{DBLP:journals/tsmc/SanfeliuF83}.  The drawback of approaches in this
category is that the graph isomorphism problem is hard to compute. The fastest
algorithm to solve it runs in quasi-polynomial time with previous solutions
being of exponential complexity \cite{DBLP:conf/stoc/Babai16}.  Considering our
method, a quasi-polynomial similarity measure is very expensive, since we have
to compute all-pairs similarity scores for a given graph dataset.\\
\shortheader{Iterative Methods:} This category of graph similarity algorithms
is based on the idea that two vertices are similar if their neighborhoods are
similar. Applying this idea iteratively over the entire graph can produce a
global similarity score when the process converges.  Based on this iterative
approach there are algorithms like \emph{SimRank} \cite{DBLP:conf/kdd/JehW02}
or the algorithm proposed by Zager et al.  \cite{DBLP:journals/appml/ZagerV08}
that compute similarities between graphs or graph nodes.  Such algorithms
compare graphs based on their topology, thus their performance depends on graph
size.  Considering our case, the efficiency of our method would depend on the
size of the graphs we compare.  Alternatively, we choose to map graphs to
feature vectors and base our similarity scores on vector comparisons. Since the
feature vectors are of low dimensionality compared to graphs, this approach is
more efficient.\\
\shortheader{Feature Vector Extraction:} These approaches are based on the idea
that similar graphs share common properties such as degree distribution,
diameter, clustering coefficient, etc.  Methods in this class represent graphs
as feature vectors. To assess the degree of similarity between graphs,
statistical tools are used to compare their feature vectors instead.  Such
methods are not as computationally demanding and thus scale better.  Drawing
from this category of measures, in our work, we base our graph similarity
computations on comparing degree distributions.\\
\shortheader{Graph Kernels:} A different approach to graph similarity comes from
the area of machine learning where kernel functions can be used to infer
knowledge about samples. A kernel can be thought of as a measure of similarity
between two objects which satisfies two mathematical properties, it is
symmetric and positive semi-definite. Graph kernels are kernel functions
constructed on graphs or graph nodes for comparing graphs or nodes
respectively. Extensive research on this area
(e.g., \cite{DBLP:journals/csr/GhoshDGQK18,DBLP:journals/sigkdd/Gartner03}) has
resulted in many kernels based on walks, paths, subgraphs, substree patterns,
etc. While computationally more expensive than feature vector extraction
similarity methods, they provide a good baseline for our modeling accuracy
evaluation.

\subsection{Graph Analytics \& Machine Learning}
Although graph analytics is a very thoroughly researched area, there exist few cases where 
machine learning techniques are used. On the subject of graph
summarization, a new approach is based on \emph{node representations} that are
learned automatically from the neighborhood of a vertex.  Such representations
are a mapping of vertices to a low-dimensional space of features that aims at
capturing their neighborhood structure.  Works in this direction
\cite{DBLP:conf/kdd/GroverL16}
focus on learning \emph{node representations} and using those for graph
summarization and/or compression. \emph{Node representations} are also
applicable in computing node or graph similarities as seen in
\cite{DBLP:conf/kdd/GroverL16} and \cite{DBLP:conf/kdd/RibeiroSF17}. However, we do not
find works employing machine learning techniques in the field of graph mining
through graph topology metric computations.  Most of the research on that field
focuses on approximation algorithms (e.g.,
\cite{DBLP:journals/datamine/RiondatoK16,DBLP:journals/jgaa/EppsteinW04}).

\section{Conclusion}

As the Graph Analytics landscape evolves, an increasing number of graph
operators are required to be executed over large graph datasets in order to
identify those of ``high interest''. To this end, we present an
operator-agnostic modeling methodology which leverages similarity between
graphs. This knowledge is used by a kNN classifier to model a given operator
allowing scientists to predict operator output for any graph without having to
actually execute the operator. We propose an intuitive, yet powerful class of
similarity measures that efficiently capture graph relations.  Our thorough
evaluation indicates that modeling a variety of graph operators is not only
possible, but it can also provide results of very high quality at considerable
speedups. This is especially true when adopting a lightweight, yet highly
effective similarity measure as those introduced by our work. Finally, our
approach appears to present similar results to state-of-the-art similarity
measures, such as \emph{D-measure}, in terms of quality, but requires orders of
magnitude less execution time.

\bibliographystyle{abbrv}
\small
\balance
\bibliography{bibliography}

\end{document}